\documentclass{article}
\usepackage{amsmath,bm,amsfonts,amssymb,setspace,graphicx,subcaption,authblk,cite}
\usepackage[margin=2cm]{geometry}
\renewcommand{\d}{\text{d}}

\begin{document}

\title{Nanothermodynamic description and molecular simulation of a single-phase fluid in a slit pore} 
\author[1,*]{Olav Galteland}
\author[1]{Dick Bedeaux}
\author[1]{Signe Kjelstrup}
\affil[1]{PoreLab, Department of Chemistry, Norwegian University of Science and Technology}
\affil[*]{olav.galteland@ntnu.no}
\date{\today}
\maketitle

\begin{abstract}
    We describe the thermodynamic state of a highly confined single-phase and single-component fluid in a slit pore using Hill's thermodynamics of small systems. This theory was more recently named nanothermodynamics. We start by constructing an ensemble of slit pores for controlled temperature, volume, surface area, and chemical potential. We present the integral and differential properties according to Hill, and use them to define the disjoining pressure. We identify all thermodynamic pressures by their mechanical counterparts in a consistent manner, and investigate the identification by molecular dynamics simulations. We define and compute the disjoining pressure, and show that it contains the standard definition. We compute the entropy and energy densities, and find in agreement with the literature, that the forces at the wall are of an energetic, not entropic nature. The subdivision potential is zero for this slit pore with large walls, but unequal to zero for related sets of control variables. We show how Hill's method can be used to find new Maxwell relations of a confined fluid, in addition to a scaling relation, which applies when the walls are separated far enough. By this expansion of nanothermodynamics, we set the stage for further developments of the thermodynamics of confined fluids, a field that is central in nanotechnology. 
\end{abstract}

\section{Introduction}

The thermodynamic state of a fluid in confinement is important for the understanding of adsorption to walls, chemical reactions, film formation and transport in porous media \cite {Israelachvili1985, Mcdonald2015, Vlugt1999, Bresme2007, Bresme2009, Galteland2020}. The molecular structuring at the walls and the forces between particles and walls are central. The change in thermodynamic properties upon confinement is substantial. This has been known for a long time \cite{Derjaguin1934,Israelachvili1985}. Derjaguin considered the measurable force that attracts or repels two walls that are close together, and defined from this the disjoining pressure \cite{Derjaguin1934}. The disjoining pressure has also been called the solvation pressure \cite{Israelachvili1985}. When the walls are far apart the disjoining pressure vanishes. It is not well known, however, how size and shape as variables affect the disjoining pressure or other properties of the confined fluid. 

Confinement is considered to be important, for instance in the context of CO$_2$ separation and sequestration by metal-organic frameworks \cite{Mcdonald2015} or for adsorption in zeolites \cite{Vlugt1999}. The disjoining pressure is of interest when studying aggregation of colloidal particles, suspended or adsorbed \cite{Bresme2007, Bresme2009, Galteland2020}. It is likely to be important also for film flow on the macroscale \cite{Moura2019}. 

More knowledge of confined fluids on the nanoscale is therefore needed. It may, for instance, help us solve the well-known up-scaling problem in porous media science \cite{Das2005}. The central problem is to understand how to integrate properties on the pore scale to the macroscale where Darcy's law applies. In order to account for shape and size effects it was recently proposed to use the four Minkowski functionals \cite{Khanamiri2018, Armstrong2019, Slotte2020}. This simplifies the description of a representative elementary volume (REV). Another procedure using the grand potential for average variables in the REV was also proposed \cite{Kjelstrup2018, Kjelstrup2019}. 

In this work we want to further examine this procedure \cite{Kjelstrup2018, Kjelstrup2019}, by looking for a way to describe the confined fluid in a pore. We are looking for a way to deal with size- and shape-dependent variables in a systematic and general manner. Two thermodynamic approaches are common. The approach following Gibbs is most popular \cite{Balbuena1993, Gubbins2014, Bedeaux2018, Gjennestad2020a, Gjennestad2020b}. But the method of Hill may offer an attractive alternative \cite{Strom2017, Galteland2019, Erdos2020, Rauter2020, Strom2020, Bedeaux2020}, partly because it may provide an independent check on Gibbs procedure, but also because general geometric scaling relations are obtainable from Hill's method. We shall see that this is also the case in the present study. 

We will pursue the method of Hill. This starts with the observation that a small system has a surface energy comparable to its bulk energy. A consequence is that the properties are not Euler homogeneous. Hill proposed to deal with such systems in an original manner \cite{Hill1963, Hill1964}. His idea was to introduce an ensemble of replicas of the small systems, on which standard thermodynamics could be applied. Hill's method has not gained much attention, in spite of a renewed effort to spur interests \cite{Hill1998, Hill2002}. 

The long-range aim of this work is to contribute to the effort of finding variables that characterize the confined fluid, for instance in a REV.  The grand potential offers one option to describe the pressure and other variables \cite{Kjelstrup2018, Kjelstrup2019}. We will pursue this route and study a single-phase and single-component fluid in a slit pore using Hill's method. The so-called integral and differential pressures introduced by Hill are central. Hill did not consider the disjoining pressure, however it is a small system property and we believe it is a concept that could benefit also from insights of nanothermodynamics. The purpose of this paper is to clarify the use of Hill's nanothermodynamics, by applying the method to a fluid in a slit pore with walls of large surface areas. This is a well studied case in the literature \cite{Israelachvili1985,Hansen1990}, and is well suited to bring out the properties of the method and its results. Grand canonical Monte Carlo and molecular dynamics simulation techniques are well suited to investigate thermodynamic relations. We will simulate a single-phase fluid in a slit pore in the grand canonical ensemble. A particular advantage of this technique is that the simulations offer a mechanical picture of the system. 

Solid-fluid and fluid-fluid interactions are considered here, but not solid-solid interactions. The solid-solid interactions will have a large effect on the thermodynamic state of the system at very small slit pore heights. We do not consider quantum effects that follow from system smallness in this work. We will consider slit pores of height $h/\lambda_{B}=6$ to $71$, where $\lambda_{B}$ is the thermal de Broglie wavelength. 

Section \ref{sec:theory} introduces the reader to Hill's nanothermodynamics. We apply this theory in section \ref{sec:Maxwell} to define size- and shape-dependent properties, and new Maxwell relations that follow from these. A definition of the disjoining pressure follows naturally in section \ref{sec:disjoining}. In order to be able to verify relations with molecular simulations as a tool, we need to identify the integral pressures and surface tension in terms the mechanical pressure tensor components. This is done in section \ref{sec:mechanical}. In section \ref{sec:method} we describe the molecular simulations. 

We proceed in section \ref{sec:results_and_discussion} to investigate relations in the theory, and illustrate them with numerical results. We compute the local mechanical and thermodynamic variables according to Hill;  \textit{i.e.} the integral and differential pressures, and the integral and differential surface tensions. The grand potential, or the replica energy, is equal to minus the integral pressure times the volume. The set of thermodynamic variables of the nanothermodynamic framework, in terms of mechanical properties, is found to be self-consistent. We offer concluding remarks in section \ref{sec:conclusion}. 

\section{Theory}
\subsection{Hill's nanothermodynamics}
\label{sec:theory}

Consider an ensemble of $\mathcal{N}$ slit pores, where each slit pore is filled with a single-phase and single-component fluid. The slit pores do not interact with each other. The $j^\text{th}$ slit pore has two parallel plane walls of area $\Omega_j$, separated by a distance $h_j$. The ensemble of slit pores has the total internal energy $U_t$, total entropy $S_t$, total volume $V_t=\sum_{j=1}^\mathcal{N} h_j\Omega_j$, total surface area $2\Omega_t=2\sum_{j=1}^\mathcal{N}\Omega_j$, and total number of particles $N$. The factor two in the total surface area arises because there are two fluid-solid surfaces of equal area per slit pore. By construction the ensemble variables $U_t, S_t, V_t, \Omega_t, N$, and $\mathcal{N}$ are Euler homogeneous functions of the first order in the number $\mathcal{N}$ of slit pores. The total differential of the total internal energy is  
\begin{equation}
	\d U_t=T\d S_t-p_\perp \d V_t+2\gamma\d \Omega_t+\mu\d  N_t+\varepsilon\d\mathcal{N}.
	\label{eq:total_internal_energy}
\end{equation}
This type of equation for the total internal energy we call the Hill-Gibbs equation \cite{Bedeaux2020}. The last term was added by Hill. The partial derivatives of the total internal energy define the temperature $T$, the normal pressure $p_\perp$, the surface tension $\gamma$ and the chemical potential $\mu$  
\begin{equation}
	\begin{split}
	    \left(\frac{\partial U_t}{\partial S_t}\right)_{V_t,\Omega_t,N_t,\mathcal{N}}& =T,
	    \qquad\left(\frac{\partial U_t}{\partial V_t}\right)_{S_t,\Omega_t,N_t,\mathcal{N}}=-p_\perp,\\
	    \left(\frac{\partial U_t}{\partial\Omega_t}\right)_{S_t,V_t,N_t,\mathcal{N}}& =2\gamma,
	    \qquad\left(\frac{\partial U_t}{\partial N_t}\right)_{S_t,V_t,\Omega_t,\mathcal{N}}=\mu.
	\end{split}
	\label{eq:partial_derivatives}
\end{equation}
The control variables in subscripts are kept constant while taking the derivatives. The volume derivative is taken while keeping the total surface area constant, which implies that the volume is changed by changing the distance between the surfaces $h_j\equiv V_j/\Omega_j$. The surface derivative is taken while keeping the volume constant, which implies that both the pore heights and surface areas are changed in such a way that the change in the total volume is zero. 

The new thermodynamic variable $\varepsilon$ is the \textit{subdivision potential}. It is defined by  
\begin{equation}
	\left(\frac{\partial U_t}{\partial\mathcal{N}}\right)_{S_t,V_t,\Omega_t,N_t}=\varepsilon.
	\label{eq:define_subdivision_potential}
\end{equation}
The subdivision potential is defined here as the increase in the total internal energy as the number of slit pores $\mathcal{N}$ increases while keeping $S_t, V_t,\Omega_t$, and $N_t$ constant. This definition is different from the definition in a previous article by us \cite{Rauter2020}, where only the entropy, volume and number of particles were kept constant and not the surface area. This led to a different expression for the subdivision potential. 

The subdivision potential is the work done on the system when adding a new slit pore while keeping the other control variables constant. The subdivision potential may be positive or negative, depending on whether work is needed or gained by adding new slit pore replicas. 

We are aiming to describe, see section \ref{sec:disjoining}, the disjoining pressure of an open slit pore when the volume, surface area, temperature and chemical potential are control variables. This set of variables is useful for describing experiments and simulations. We use the average volume per slit pore $V=V_t/\mathcal{N}$ and the average surface area per slit pore $2\Omega =2\Omega_t/\mathcal{N}$, rather than the total volume $V_t$ and the total surface area $2\Omega_t$. In order to obtain an appropriate Hill-Gibbs equation for this case, we substitute the total volume with the average volume $V_t=V\mathcal{N}$ and total surface area with the average surface area $2\Omega_t=2\Omega\mathcal{N}$. The total differentials of the volume and surface area are  
\begin{equation}
	\d(V\mathcal{N})=\mathcal{N}\d V+V\d\mathcal{N}\quad\text{and}\quad\d(\Omega \mathcal{N)}=\mathcal{N}\d\Omega +\Omega\d\mathcal{N}.
\end{equation}
By introducing this into the Hill-Gibbs equation, see equation \ref{eq:total_internal_energy}, we obtain  
\begin{equation}
	\d U_t=T\d S_t-p_\perp \mathcal{N}\d V+2\gamma\mathcal{N}\d \Omega +\mu \d N_t+(\varepsilon-p_\perp V+2\gamma\Omega)\d \mathcal{N}.
	\label{eq:total_internal_energy2}
\end{equation}
The parenthesis define the replica energy  
\begin{equation}
	X(T,V,\Omega,\mu)\equiv\varepsilon-p_\perp V+2\gamma\Omega.
	\label{eq:grand_potential}
\end{equation}
The subdivision potential $\varepsilon$, normal pressure $p_\perp$, and surface tension $\gamma$ depend on the control variable set $T,V,\Omega$, and $\mu$. The replica energy density will be used to define the disjoining pressure. 

The grand partition function covers all microstates available to the slit pore. In order to calculate this partition function one chooses a volume $V$ and a surface area $\Omega$. Both can be varied independently. For large $h$ and $\Omega$ only a change in the volume $V=h\Omega$, and not in $h$ and $\Omega$ separately, matters. For a small volume it is necessary to use the volume $V$ and the surface area $\Omega$ as independent variables. This is because the same volume change due to a change of the pore height $h=V/\Omega$ or due to a change of the surface area $\Omega$ produces different changes in the partition function, and therefore in the thermodynamic variables. 

The replica energy was identified by Hill as the grand potential, here equal to minus the integral pressure times the volume \cite{Hill1963},  
\begin{equation}
	X(T,V,\Omega,\mu)=-\hat{p}V=-\hat{p}_\perp V+2\hat{\gamma}\Omega.
	\label{eq:integral_pressure}
\end{equation}
In the last equality, we have chosen to identify the integral pressure as the integral normal pressure times the volume minus the integral surface tension times the surface area. We will refer to $\hat{p}_\perp$, $\hat{\gamma}$ and $\hat{p}$ as the integral normal pressure, the integral surface tension and the integral pressure, respectively, while $p_\perp$ and $\gamma$ are the differential normal pressure and the differential surface tension, respectively. The names integral and differential pressure where coined by Hill to reflect that the differential pressure involves the differential of the integral pressure. We have chosen a control variable set with volume and surface area, such that we do not have a differential pressure but a differential normal pressure $p_\perp$ and differential surface tension $\gamma$ in its place. 

From equations \ref{eq:grand_potential} and \ref{eq:integral_pressure} it follows that the subdivision potential is  
\begin{equation}
	\varepsilon =\left( p_\perp -\hat{p}_\perp \right) V-2\left(\gamma-\hat{\gamma}\right)\Omega.
	\label{eq:epsilon}
\end{equation}
The subdivision potential $\varepsilon$ indicate that the integral normal pressure and the integral surface tension may be different from the corresponding differential variables. 

\subsection{Maxwell relations for a slit pore}
\label{sec:Maxwell}

Using that the total internal energy is Euler homogeneous of the first order in the number of slit pores, see equation \ref{eq:total_internal_energy2}, it follows that the total internal energy is  
\begin{equation}
	U_t=TS_t+\mu N_t+X\mathcal{N}.
	\label{eq:total_internal_energy3}
\end{equation}
We introduce the average internal energy, entropy and number of particles per slit pore  
\begin{equation}
	U_t=U\mathcal{N},\qquad S_t=S\mathcal{N},\quad\text{and}\quad N_t=N\mathcal{N}.
	\label{eq:average_properties}
\end{equation}
By introducing the average properties into the total internal energy in equation \ref{eq:total_internal_energy3} we obtain the internal energy per slit pore  
\begin{equation}
	U=TS+\mu N+(\varepsilon-p_\perp V+2\gamma\Omega)=TS+\mu N+X.
	\label{eq:internal_energy}
\end{equation}
Substituting the averages properties per slit pore in equation \ref{eq:average_properties} into the corresponding Hill-Gibbs equation, see equation \ref{eq:total_internal_energy2}, and using the internal energy in equation \ref{eq:internal_energy}, we obtain the total differential of the internal energy  
\begin{equation}
	\d U=T\d S-p_\perp \d V+2\gamma\d\Omega +\mu\d N.
	\label{eq:internal_energy2}
\end{equation}
By differentiating the internal energy in equation \ref{eq:internal_energy} and using the total differential of the internal energy in equation \ref{eq:internal_energy2}, we obtain the total differential of the replica energy  
\begin{equation}
	\d X=-\d(\hat{p}_\perp V-2\hat{\gamma}\Omega)=-\d(\hat{p}V)=-S\d T-p_\perp \d V+2\gamma\d\Omega-N\d\mu.
	\label{eq:hill_gibbs_duhem}
\end{equation}
This equation was termed the Hill-Gibbs-Duhem equation \cite{Bedeaux2020}, because it reduces to the Gibbs-Duhem equation for a large system. It follows that the partial derivatives of the grand potential is  
\begin{equation}
    \begin{split}
	\left(\frac{\partial\left(\hat{p}V\right)}{\partial T}\right)_{V,\Omega,\mu}& =-\left(\frac{\partial X}{\partial T}\right)_{V,\Omega,\mu}=S,\\
	\left(\frac{\partial\left(\hat{p}V\right)}{\partial V}\right)_{T,\Omega,\mu}& =-\left(\frac{\partial X}{\partial V}\right)_{T,\Omega,\mu}=p_\perp ,\\
	\left(\frac{\partial\left(\hat{p}V\right)}{\partial\Omega}\right)_{T,V,\mu}& =-\left(\frac{\partial X}{\partial\Omega}\right)_{T,V,\mu}=-2\gamma,\\
	\left(\frac{\partial\left(\hat{p}V\right)}{\partial\mu}\right)_{T,V,\Omega}& =-\left(\frac{\partial X}{\partial\mu}\right)_{T,V,\Omega}=N.
	\label{eq:partial_derivatives2}
    \end{split}
\end{equation}
Rather than the names replica energy or grand potential, we name from now $X=-\hat{p}V$ by minus the integral pressure times the volume. In section \ref{sec:mechanical} we will define the integral pressure in terms of the average tangential mechanical pressure which can be calculated from molecular simulations. In the simulations we consider surface areas $\Omega$ much larger than the diameter of the fluid particles and heights $h=V/\Omega$ comparable to the diameter of the fluid particles. This implies that $\hat{p}$, $\hat{p}_\perp$, $\hat{\gamma}$, $u\equiv U/V$, $s\equiv S/V$, $p_\perp$, $\gamma$, and $\rho\equiv N/V$ do not depend on $\Omega$, but the variables will depend on the height $h$ and therefore on the volume $V$. 

The volume and surface derivative can be rewritten in terms of derivatives of the slit pore height and surface area. These derivatives are needed to calculate the differential normal pressure and differential surface tension. The differential normal pressure is  
\begin{equation}
	p_\perp =\left(\frac{\partial(\hat{p}V)}{\partial V}\right)_{T,\Omega,\mu}=\hat{p}+V\left(\frac{\partial\hat{p}}{\partial V}\right)_{T,\Omega,\mu}=\hat{p}+h\left(\frac{\partial\hat{p}}{\partial h}\right)_{T,\Omega,\mu}.
	\label{eq:differential_normal_pressure}
\end{equation}
The differential surface tension is  
\begin{equation}
	\gamma =-\frac{V}{2}\left(\frac{\partial\hat{p}}{\partial\Omega}\right)_{T,V,\mu}=\frac{h^{2}}{2}\left(\frac{\partial\hat{p}}{\partial h}\right)_{T,\Omega,\mu}-\frac{V}{2}\left(\frac{\partial\hat{p}}{\partial\Omega}\right)_{T,h,\mu}.
	\label{eq:differential_surface_tension}
\end{equation}
Combining the equations for the differential normal pressure and surface tension it follows that the integral pressure is  
\begin{equation}
	\hat{p}=p_\perp -\frac{2}{h}\gamma-\Omega\left(\frac{\partial\hat{p}}{\partial\Omega}\right)_{T,\mu,h}.
	\label{eq:integral_pressure0}
\end{equation}
The last term is the equal to minus the subdivision potential divided by volume, see equations \ref{eq:integral_pressure} and \ref{eq:epsilon}. The subdivision potential is  
\begin{equation}
	\varepsilon =\Omega V\left(\frac{\partial\hat{p}}{\partial\Omega}\right)_{T,\mu,h}.
	\label{eq:subdivision_zero}
\end{equation}
This implies that in the grand canonical ensemble with $T,V,\Omega,\mu$ as control variables, the thermodynamic description of the slit pore with a small height is the same as for the slit pore with a large height. However, it changes when the integral pressure depends on the surface area. This may not be the case for any other set of control variables. In general the properties of a small system (confined fluid) depend on the set of control variables.  

In this work we deal with large surface areas, such that the integral pressure does not depend on it. As a consequence the subdivision potential is zero. We will find that the integral and differential normal pressure are equal. Using that $\varepsilon =0$ it follows that the integral and differential surface tensions are also equal for a large surface area, see equation \ref{eq:epsilon},  
\begin{equation}
	\hat{p}_\perp =p_\perp \quad\text{and}\quad\hat{\gamma}=\gamma.
	\label{eq:int_diff_equal}
\end{equation}
We shall investigate the first identity for large surface areas using molecular simulations.  

From equation \ref{eq:hill_gibbs_duhem} we obtain the following Maxwell relations of the differential surface tension,  
\begin{equation}
	\begin{split}
	\left(\frac{\partial\gamma}{\partial T}\right)_{V,\Omega,\mu}& =-\frac{ 1}{2}\left(\frac{\partial S}{\partial\Omega}\right)_{T,\mu,V}=\frac{ h^{2}}{2}\left(\frac{\partial s}{\partial h}\right)_{T,\mu,\Omega},\\
	\left(\frac{\partial\gamma}{\partial\mu}\right)_{T,V,\Omega}& =-\frac{ 1}{2}\left(\frac{\partial N}{\partial\Omega}\right)_{T,\mu,V}=\frac{ h^{2}}{2}\left(\frac{\partial\rho}{\partial h}\right)_{T,\mu,\Omega},\\
	\left(\frac{\partial\gamma}{\partial V}\right)_{T,\Omega,\mu}& =-\frac{ 1}{2}\left(\frac{\partial p_\perp }{\partial\Omega}\right)_{T,\mu,V}=\frac{h}{2\Omega}\left(\frac{\partial p_\perp }{\partial h}\right)_{T,\mu,\Omega}.
	\end{split}
\end{equation}
In the second identity we have used that the entropy density, fluid number density and differential normal pressure are independent of the surface area for large surface areas. Other Maxwell relations are possible, see the Hill-Gibbs-Duhem equation \ref{eq:hill_gibbs_duhem}. The last equality can be written as  
\begin{equation}
	\left(\frac{\partial\gamma}{\partial h}\right)_{T,\mu}=\frac{h}{2}\left(\frac{\partial p_\perp }{\partial h}\right)_{T,\mu,\Omega}.
	\label{23}
\end{equation}
The important implication of this expression is that the constant nature of one variable implies the constant nature of the other for large pore heights. With the mechanical description of the integral properties given in section \ref{sec:mechanical} all the above thermodynamic relations can be tested. 

\subsection{The disjoining pressure}
\label{sec:disjoining}

We define the excess replica energy as the replica energy of a slit pore of height $h$ minus the replica energy of the slit pore where the slit pore height approaches infinity,  
\begin{equation}
	X^\text{ex}\ =X-\lim_{h\rightarrow\infty}X.
\end{equation}
We denote the thermodynamic variables where the slit pore height approaches infinity with $\infty$ in superscript,  
\begin{equation}
	\lim_{h\rightarrow\infty}X=X^\infty.
\end{equation}
The excess replica energy can be written in terms of the excess integral pressure $\hat{p}^\text{ex}$, excess integral normal pressure $\hat{p}_\perp^\text{ex}$, and excess integral surface tensions $\hat{\gamma}^\text{ex}$ 
\begin{equation}
	X^\text{ex}\ =-\hat{p}^\text{ex}\ V=-\hat{p}_\perp ^\text{ex}V+2\hat{\gamma}^\text{ex}\Omega.
\end{equation}
We define now the disjoining pressure as the excess normal pressure,  
\begin{equation}
	\Pi(h)\equiv\hat{p}_\perp ^\text{ex}\equiv\hat{p}_\perp -p_\perp ^\infty.
	\label{eq:def_disjoining}
\end{equation}
Other possible names are minus the excess replica density, or the excess grand potential density. 

This definition of the disjoining pressure is different from the typical definition found in the literature \cite{Derjaguin1934, Hansen1990, Radke2015}. Typically the disjoining pressure is defined to be equal to what we in this work call the excess differential normal pressure, where the excess is relative to a bulk fluid. For a large surface area our molecular dynamics simulations find that $\hat{p}_\perp=p_\perp$. We will furthermore show that the integral normal pressure, as the slit pore height approaches infinite separation, equals the bulk fluid pressure $p_\perp ^\infty=p^{b}$. This shows that our choice is equivalent to the usual definition. 

\subsection{A mechanical description of the slit pore}
\label{sec:mechanical}

The thermodynamic description of integral and differential pressures and surface tensions has its mechanical equivalent description in terms of components of the mechanical pressure tensor. A recent discussion on this topic clarified the challenge of translating the mechanical pressure tensor into a thermodynamic scalar variable in a meaningful manner \cite{Long2011, Dijk2020, Long2020}. We will identify the integral normal pressure, integral surface tension, and integral pressure in such a way that thermodynamic framework is self-consistent. However, we do not claim that this is the only valid choices of thermodynamic pressures and tensions in terms of the mechanical pressure tensor. 

The mechanical pressure tensor of a heterogeneous system is ambiguously defined. This has been known for a long time, at least since the work by Irving and Kirkwood in the 1940s. It was shown by Schofield and Henderson  \cite{Schofield1982} that the ambiguity is due to the arbitrary choice of the integration contour $C_{ij}$, which is needed to calculate the configurational contribution to the pressure tensor. The local mechanical pressure tensor is calculated in a subvolume $V_l$ as a sum of the kinetic and the configurational contributions,  
\begin{equation}
	P_{\alpha\beta}(x)=P_{\alpha\beta}^{k}+P_{\alpha\beta}^{c}.
	\label{eq:mechanical_pressure_tensor2}
\end{equation}
Upper case $P$ is used to denote mechanical pressure tensor components in order to distinguish them from the thermodynamic pressures, which are denoted by lower case $p$. The kinetic contribution is the ideal contribution to the mechanical pressure and is calculated as  
\begin{equation}
	P_{\alpha\beta}^{k}=\frac{1}{V_l}\left\langle\sum_{i\in V_l}m_{i}v_{i,\alpha}v_{i,\beta}\right\rangle.
	\label{eq:kinetic_pressure}
\end{equation}
The sum with subscript $i\in V_l$ represents a sum over all particles in the subvolume $V_l$. The particle mass is $m_{i}$ and $v_{i,\alpha}$ is the velocity in the $\alpha$-direction. The solid walls do not have a velocity and consequently do not contribute to the kinetic pressure. The brackets $\langle\dots\rangle$ represent the grand canonical ensemble average. The
configurational contribution is the non-ideal contribution to the mechanical pressure and is calculated as  
\begin{equation}
	P_{\alpha\beta}^{c}=-\frac{1}{2V_l}\left\langle\sum_{i=1}^{N}\sum_{\substack{ j=1\\
	 j\neq i}}^{N}f_{ij,\alpha}\int_{C_{ij}\in V_l}\d l_{\beta}\right\rangle.
	\label{eq:conf_pressure}
\end{equation}
The sums represents a sum over all particle pairs. The $\alpha$-component of the force vector acting on particle $i$ due to particle $j$ is $f_{ij,\alpha}$. The fluid-fluid and fluid-solid interactions contribute to the configurational pressure, the solid-solid interaction is zero and does not contribute to the pressure. The line integral is the $\beta$-component of the part of the contour $C_{ij}$ contained in the subvolume $V_l$. 

The contour $C_{ij}$ is the source of the ambiguity of the mechanical pressure tensor, it can be any continuous line from the centers of particles $i$ to $j$. The Harasima \cite{Harasima1958} and the Irving-Kirkwood \cite{Irving1950} contours are two common choices for $C_{ij}$. The Harasima contour is defined as two continuous line segments, a line from the center of particle $i$ parallel to the surface and a line normal to the surface to the center of particle $j$. The Irving-Kirkwood is the straight line from particle $i$ to $j$. For flat surfaces they are equal. However for spherical surfaces the Harasima contour does not obey momentum balance \cite{Hafskjold2002}. There are cases where the Harasima contour is more useful than the Irving-Kirkwood contour \cite{Shi2020}. In this work we will use the equations by Ikeshoji \textit{et al.} \cite{Ikeshoji2003} with the Irving-Kirkwoood contour to calculate the mechanical pressure tensor. 

We will only consider surface areas $\Omega$ much larger in both directions than the fluid particle diameter. This implies the mechanical pressure tensor does not depend on $\Omega$. It does however depend on the height $h\equiv V/\Omega$ and therefore on the volume $V$. 

Using the translational symmetry of the slit pore in the $y$- and $z$-direction, where the $x$-direction is normal to the solid surface, the equilibrium mechanical pressure tensor in the slit pore has the form  
\begin{equation}
	\mathbf{P}(x;h)=P_\perp (h)\mathbf{e}_{x}\mathbf{e}_{x}+P_{\parallel}(x;h)(\mathbf{e}_{y}\mathbf{e}_{y}+\mathbf{e}_{z}\mathbf{e}_{z}).
	\label{eq:mechanical_pressure_tensor}
\end{equation}
Where $\mathbf{e}_{x}$, $\mathbf{e}_{y}$ and $\mathbf{e}_{z}$ are the unit vectors in the $x$-, $y$- and $z$-directions. The normal pressure tensor component is equal to the $xx$-component, and tangential pressure tensor component is the average of the $yy$- and $zz$-components,  
\begin{equation}
	P_\perp (h)=P_{xx}\quad\text{and}\quad P_{\parallel}(x,h)=\frac{1}{2}\left( P_{yy}+P_{zz}\right).
	\label{4.3}
\end{equation}
Mechanical equilibrium requires that the tangential pressure is independent of the $y$- and $z$-coordinates, but depends on the $x$-coordinate. The normal pressure is independent of all spatial coordinates. 

We identify the thermodynamic integral normal pressure in terms of the volume integral of the normal mechanical pressure divided by the volume. However, since the normal mechanical pressure and the area are constant everywhere, this simplifies to  
\begin{equation}
	\hat{p}_\perp (h)\equiv\frac{1}{h}\int_{0}^{h}P_\perp (h)x=P_\perp (h).
	\label{eq:def_integral_normal_pressure}
\end{equation}
The integral normal pressure in a large pore is equal to the pressure in a bulk phase in equilibrium with the pore. The difference in the integral normal pressure in a liquid and vapor phase in a slit pore is described the Young-Laplace equation, while the integral pressure is the same in both the liquid and vapor phase \cite{Rauter2020}. An alternative route to the integral normal pressure is via the local fluid number density profile \cite{Evans1986}  
\begin{equation}
	P_\perp (h)=\int_{0}^{h}f_{fs}(x)\rho(x,h)x,
	\label{eq:disjoining_normal_density_profile}
\end{equation}
where $f_{fs}(x)$ is the fluid-solid force and $\rho(x,h)$ is the local fluid number density. The fluid density and fluid-solid force are uniquely defined, and do not have the inherent problem that the mechanical pressure tensor has. We can use this equation to validate our method of calculating the integral normal pressure. 

We identify further the integral surface tension as the integral of the normal minus tangential pressure tensor components,  
\begin{equation}
	\hat{\gamma}(h)\equiv\frac{1}{2}\int_{0}^{h}\left( P_\perp (h)-P_{\parallel}(x;h)\right)\d x,
	\label{eq:def_integral_surface_tension}
\end{equation}
where the factor half is due to the fact that there are two fluid-solid surfaces. It follows from equation \ref{eq:integral_pressure} together with equations \ref{eq:def_integral_normal_pressure} and \ref{eq:def_integral_surface_tension} that the integral pressure is,  
\begin{equation}
	\hat{p}(h)=\frac{1}{h}\int_{0}^{h}P_{\parallel}(x;h)\d x.
	\label{eq:def_integral_pressure}
\end{equation}
As shown by Harasima \cite{Harasima1958} and Schofield and Henderson \cite{Schofield1982} a sufficiently large volume integral of the mechanical pressure tensor components does not depend on the choice of the integral contour $C_{ij}$. We have identified the integral normal pressure, integral surface tension and integral pressure in terms of the mechanical pressure tensor. The local mechanical pressure tensor has an inherent problem, specifically that the contour $C_{ij}$ can be any continuous line from $i$ to $j$. However the thermodynamic variables do not have this inherent problem. This is because we integrate the the local mechanical pressure tensor across the whole volume $V$. This volume integral includes all interactions. 

The internal energy can be calculated as the sum of the kinetic and potential energy,  
\begin{equation}
	U=E_{k}+E_{p}.
	\label{eq:internal_energy3}
\end{equation}
By dividing the internal energy in equation \ref{eq:internal_energy} by the volume we obtain the internal energy density $u=U/V$. We also use the entropy density $s=S/V$, and fluid number density $\rho =N/V$. By rearranging the equation we obtain the entropy density as  
\begin{equation}
	s=\frac{1}{T}\left(u+\hat{p}-\mu\rho\right),
	\label{eq:entropy}
\end{equation}
where the internal energy density, integral pressure and fluid number density are known. The volume, chemical potential and temperature are imposed on the system. 

\section{Simulation details}
\label{sec:method}

The thermodynamic state of slit pores of varying heights $h$ was investigated by using grand-canonical Monte Carlo (GCMC) \cite{Frenkel2001} in combination with molecular dynamic (MD) simulations with the Nosé-Hoover thermostat \cite{Shinoda2004}. This produced the grand
canonical ensemble, \textit{i.e.} constant chemical potential, temperature, volume and surface area. The GCMC method inserted and deleted fluid particles to and from the simulation box from an imaginary fluid particle reservoir at the same temperature and chemical potential. This controlled the chemical potential of the fluid in the slit pore. The MD procedure updated the positions and velocities of the fluid particles and controlled the temperature with the Nosé-Hoover thermostat. 

The simulations were carried out using LAMMPS \cite{Plimpton1995}. The local mechanical pressure tensor was calculated by post-processing the particle trajectories with in-house software. The chemical potential and temperature were kept constant at $\mu ^{\ast}=1$ and $T^{\ast}=2$. The critical temperature of the Lennard-Jones/spline fluid is $T_{c}^{\ast}=0.855$ \cite{Hafskjold2019}. All units in this work are in reduced Lennard-Jones units, see table \ref{tab:units} for a definition. 

\begin{table}
    \caption{The reduced units are denoted with an asterisk in superscript, for example $T^{\ast}$. The variables are reduced using the molecular diameter $\sigma $, potential well depth $\epsilon $, fluid particle mass $m$ and Boltzmann constant $k_{\text{B}}$} 
    \label{tab:units}
    \centering  
    \begin{tabular}{cc}
        \hline
        Description & Definition  \\ 
        \hline 
        Energy & $E^{\ast}=E/\epsilon $  \\  
        Entropy & $S^{\ast}=S/k_\text{B}$  \\  
        Temperature & $T^{\ast}=Tk_\text{B}/\epsilon $  \\  
        Distance & $x^{\ast}=x/\sigma $  \\  
        Pressure & $p^{\ast}=p\sigma ^{3}/\epsilon $  \\  
        Chemical potential & $\mu ^{\ast}=\mu /\epsilon $  \\ 
        \hline 
    \end{tabular}%
\end{table}

The simulation box was a rectangular cuboid of side lengths $L_{x},L_{y}=L_{z}$. The side lengths $L_{y}=L_{z}$ were chosen such that the surface area was large. Large in this context indicates large enough for $\hat{p}$, $\hat{p}_\perp $, $\hat{\gamma}$, $p_\perp $, and $\gamma$ to be independent of the surface area $\Omega = L_{y}L_{z}$. The simulation box size was decided such that the average number of fluid particles was approximately $2\times 10^{4}$. The simulation box was periodic in the $y$- and $z$-directions, and non-periodic in the $x$-direction. This implies that the particles did not interact across the simulation box boundary in the $x$-direction. 

The fluid-fluid and fluid-solid interaction was modeled with the Lennard-Jones/spline potential \cite{Hafskjold2019}. The fluid-fluid and fluid-solid interactions were equal. The potential energy of a fluid-fluid or fluid-solid pair separated by a distance $r$ was  
\begin{equation}
	u^{LJ/s}(r)= 
	\begin{cases} 
	    4\epsilon\left[\left(\frac{\sigma}{r}\right) ^{12}-\left(\frac{\sigma}{r}\right) ^{6}\right] &\text{if } r<r_{s}\\
	    a(r-r_{c})^{2}+b(r-r_{c})^{3} &\text{if } r_{s}<r<r_{c}\\
	    0 &\text{else},
	\end{cases}
	\label{eq:ljs}
\end{equation}
where $r_{s}$, $a$, $b$ and $r_{c}$ were chosen such that the potential energy and the force were continuous at the inflection point $r=r_{s}$ and the cut-off $r=r_{c}$. The solid walls were placed at the simulation box boundaries $x=-L_x/2$ and $x=L_x/2$. The distance between the fluid and solids were $r=|x_f-L_x/2|$ and $r=|x_f+L_x/2|$, where $x_f$ is the $x$-position of the fluid particle. 

\begin{figure}
    \centering
    \includegraphics[width=0.6\textwidth]{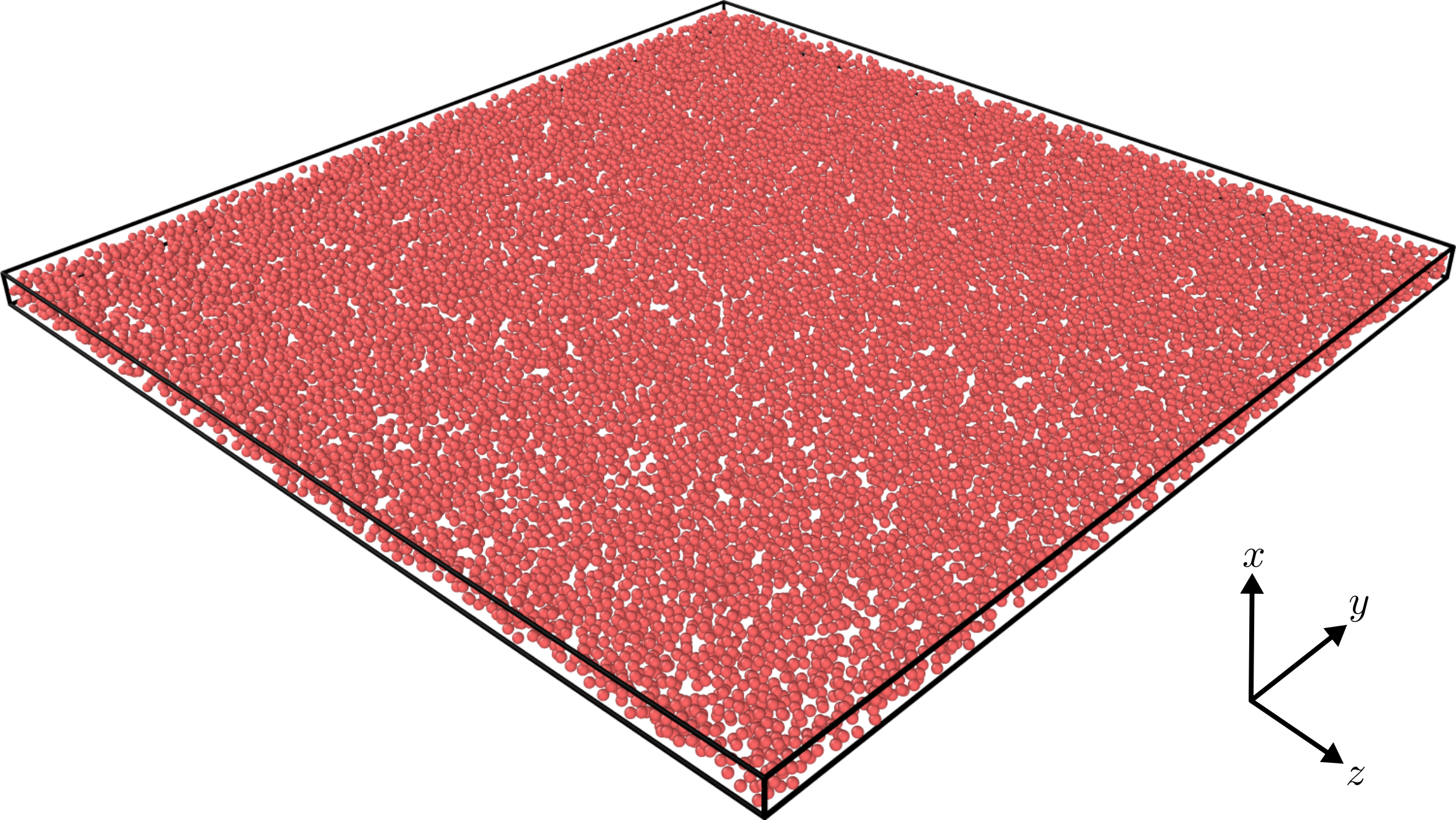}  
    \caption{Visualization of the fluid particles in a slit pore of height $L_x=4\sigma$, chemical potential $\mu ^{\ast}=1$, and temperature $T^{\ast}=2$. The fluid particles are rendered in red, and their diameter is rendered at $\sigma$. The solid is not rendered. The solid lines illustrates the edges of the simulation box. The simulation was rendered with OVITO \cite{Stukowski2009}.} 
    \label{fig:system_h4}
\end{figure}

The dividing surfaces of the fluid-solid surfaces were chosen to be at $x=-L_x/2$ and $x=L_x/2$. The slit pore height was consequently determined to be $h = L_{x}$. Other choices of the dividing surface are possible, for example the Gibbs dividing surface or the surface of tension. When $L_x<2r_c$ the fluid particle can interact with both solid walls. See figure \ref{fig:system_h4} for an visualization of the simulation box for the case $L_x=4\sigma$. 

The mechanical pressure tensor was calculated in thin rectangular cuboids, called layers $l$, of side lengths $\Delta x,L_{y},L_{z}$. The thickness of the layers was $\Delta x=0.005\sigma $ and the number of layers was $n_l=h/\Delta x$. The diagonal components of the mechanical pressure tensor was calculated using equations \ref{eq:mechanical_pressure_tensor}, \ref{eq:kinetic_pressure} and \ref{eq:conf_pressure}. 

The kinetic energy was calculated as the sum of the kinetic energy for each fluid particle and the potential energy was calculated as the sum of the potential energy of each fluid-fluid and fluid-solid pairs,  
\begin{equation}
	E_{k}=\frac{1}{2}\sum_{i=1}^{N}m_{i}(\mathbf{v}_{i}\cdot\mathbf{v}_{i})\quad\text{and}\quad E_{p}=\sum_{i=1}^{N}\sum_{j>i}^{N}u^{LJ/s}(r_{ij}).
\end{equation}
The sum of the kinetic and potential energies were used to calculate the internal energy and entropy densities. 

\section{Results and discussion}
\label{sec:results_and_discussion}

\begin{figure}
    \centering 
    \begin{subfigure}{0.5\textwidth}
        \includegraphics[width=\textwidth]{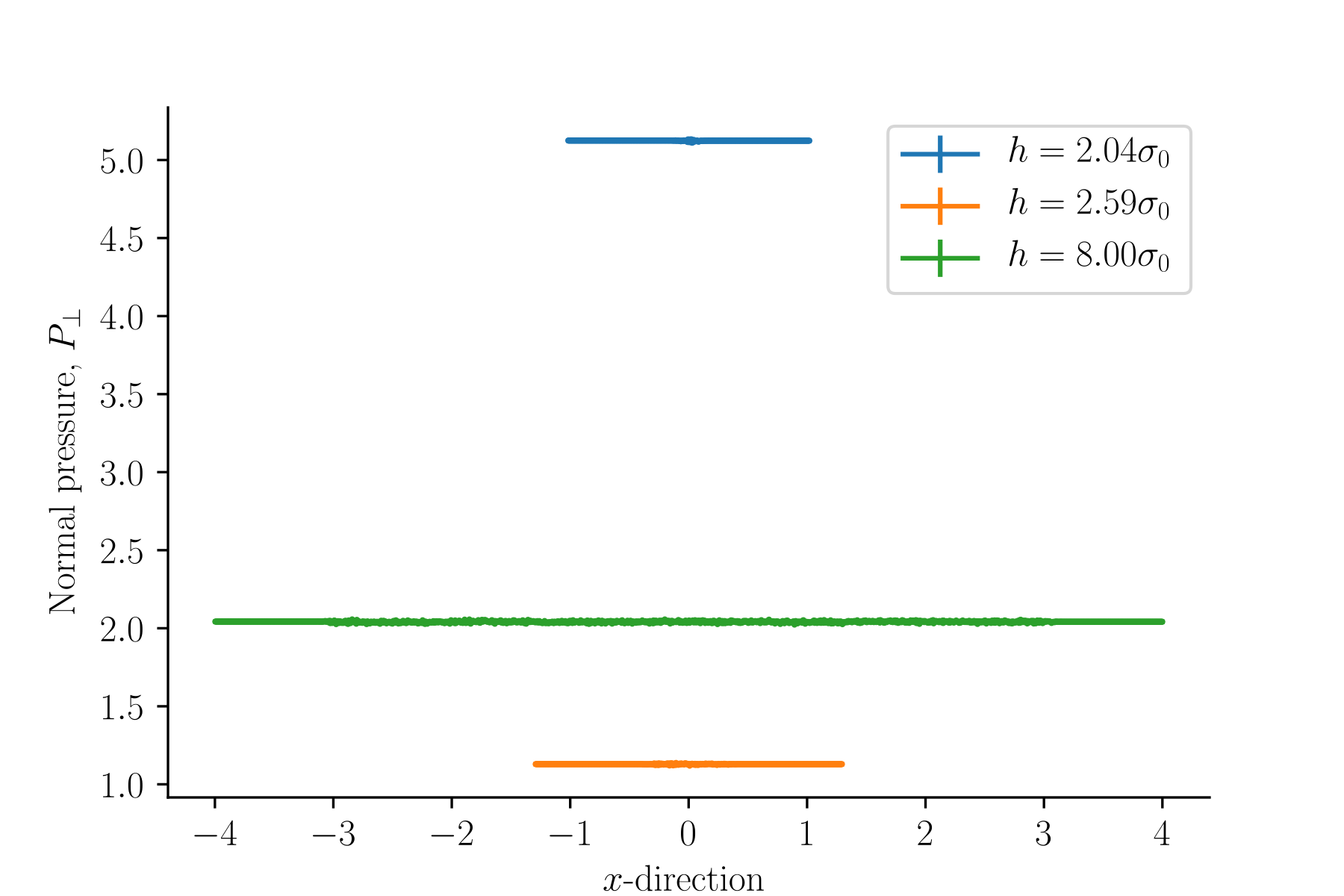}          
        \caption{}                
        \label{fig:profile_pn}        
    \end{subfigure} 
    \begin{subfigure}{0.5\textwidth}
        \includegraphics[width=\textwidth]{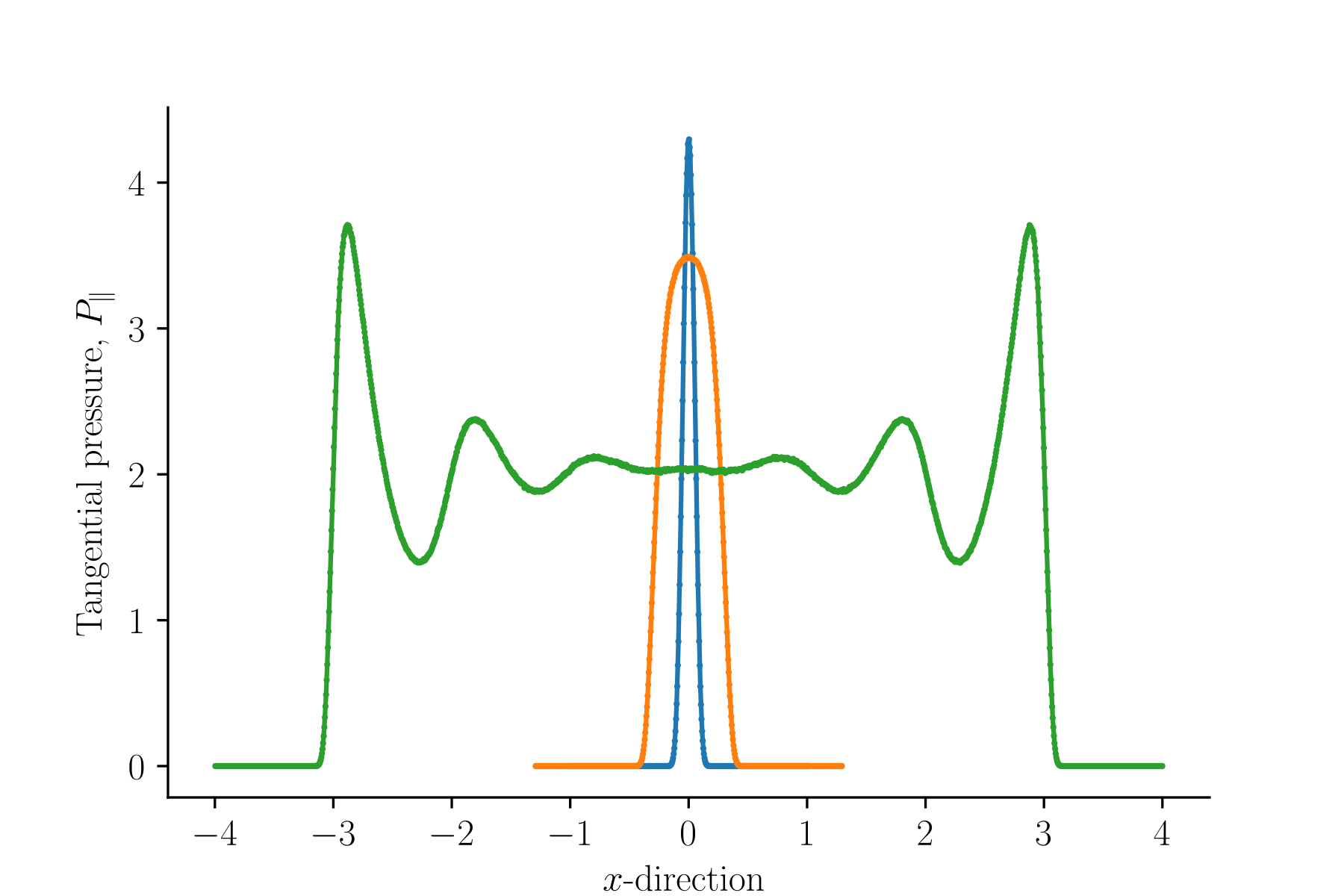}         
        \caption{}                
        \label{fig:profile_pt}        
    \end{subfigure}  
    \begin{subfigure}{0.5\textwidth}
        \includegraphics[width=\textwidth]{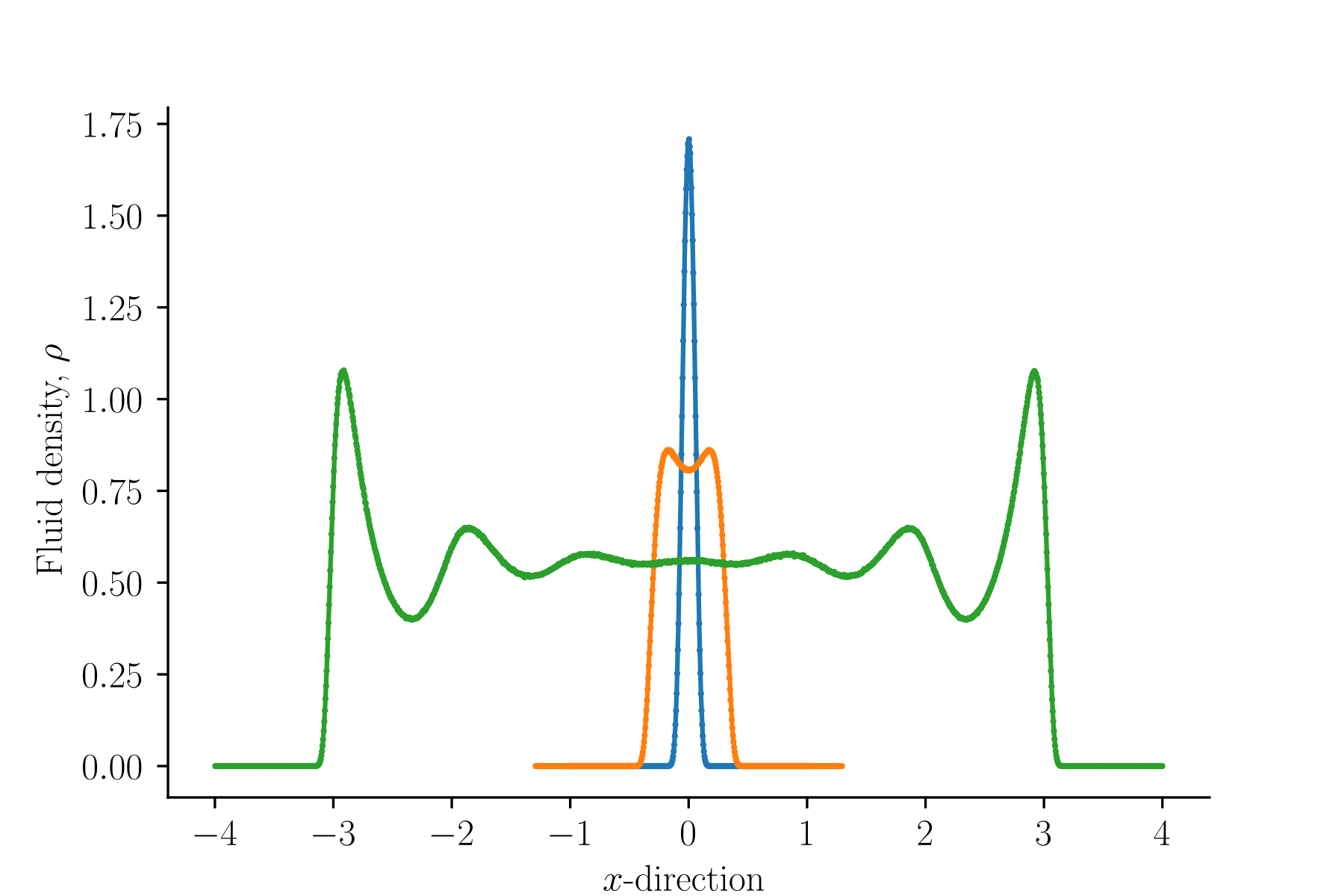}                
        \caption{}                
        \label{fig:profile_density}        
    \end{subfigure}  
    \caption{\textbf{(a)} Normal mechanical pressure, \textbf{(b)} tangential mechanical pressure, and \textbf{(c)} fluid number density $\rho$ as a function of the $x$-direction for slit pore heights $h=2.04\sigma$, $2.59\sigma$, and $8\sigma$.}
    \label{fig:profiles}
\end{figure}

\begin{figure}
    \centering
    \includegraphics[width=0.8\textwidth]{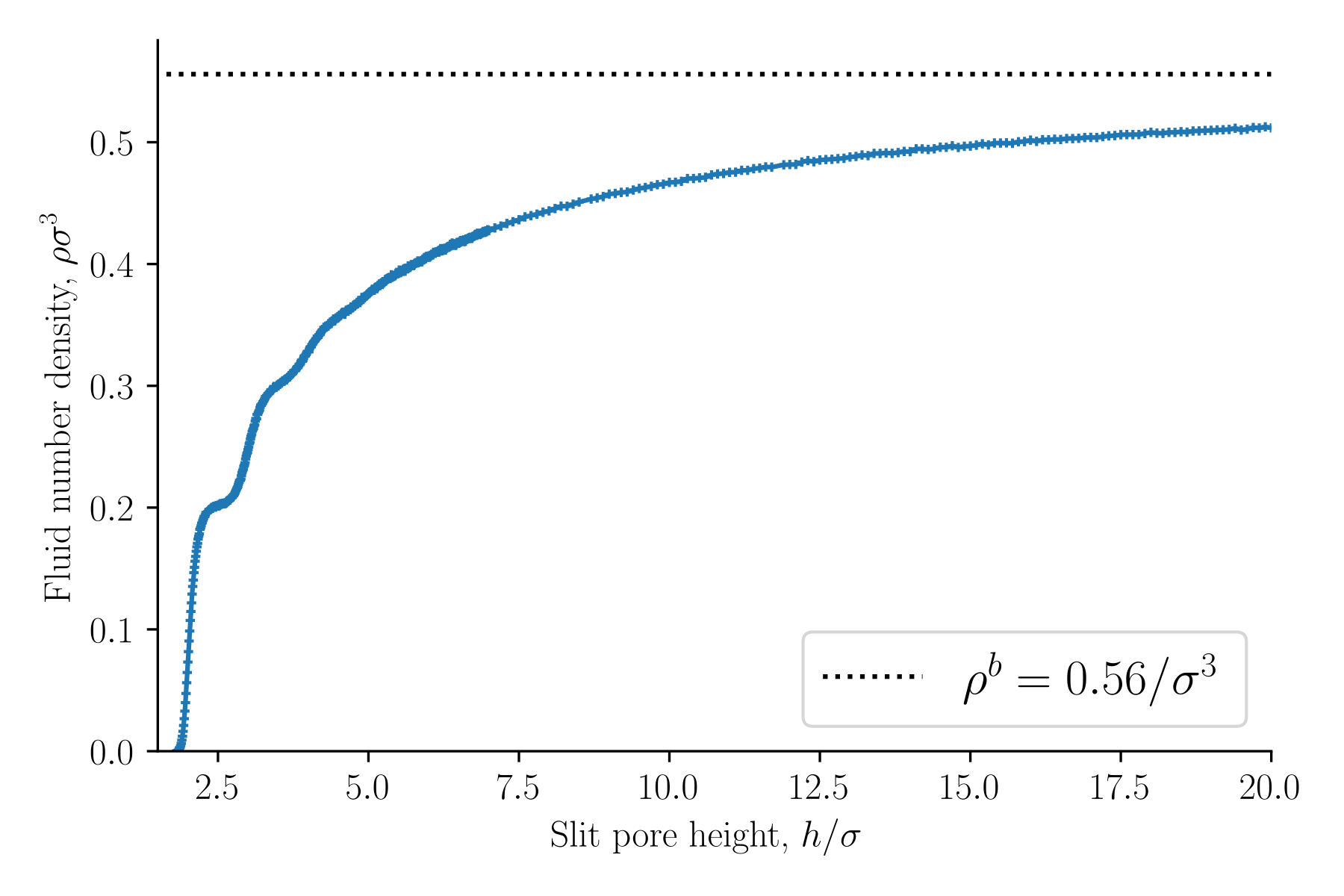}  
    \caption{Fluid number density $\rho=N/V$ as a function of slit pore
    height $h$. The bulk fluid number density is shown as a dashed line.} \label{fig:density}
\end{figure}

\begin{figure}
    \centering
    \includegraphics[width=0.8\textwidth]{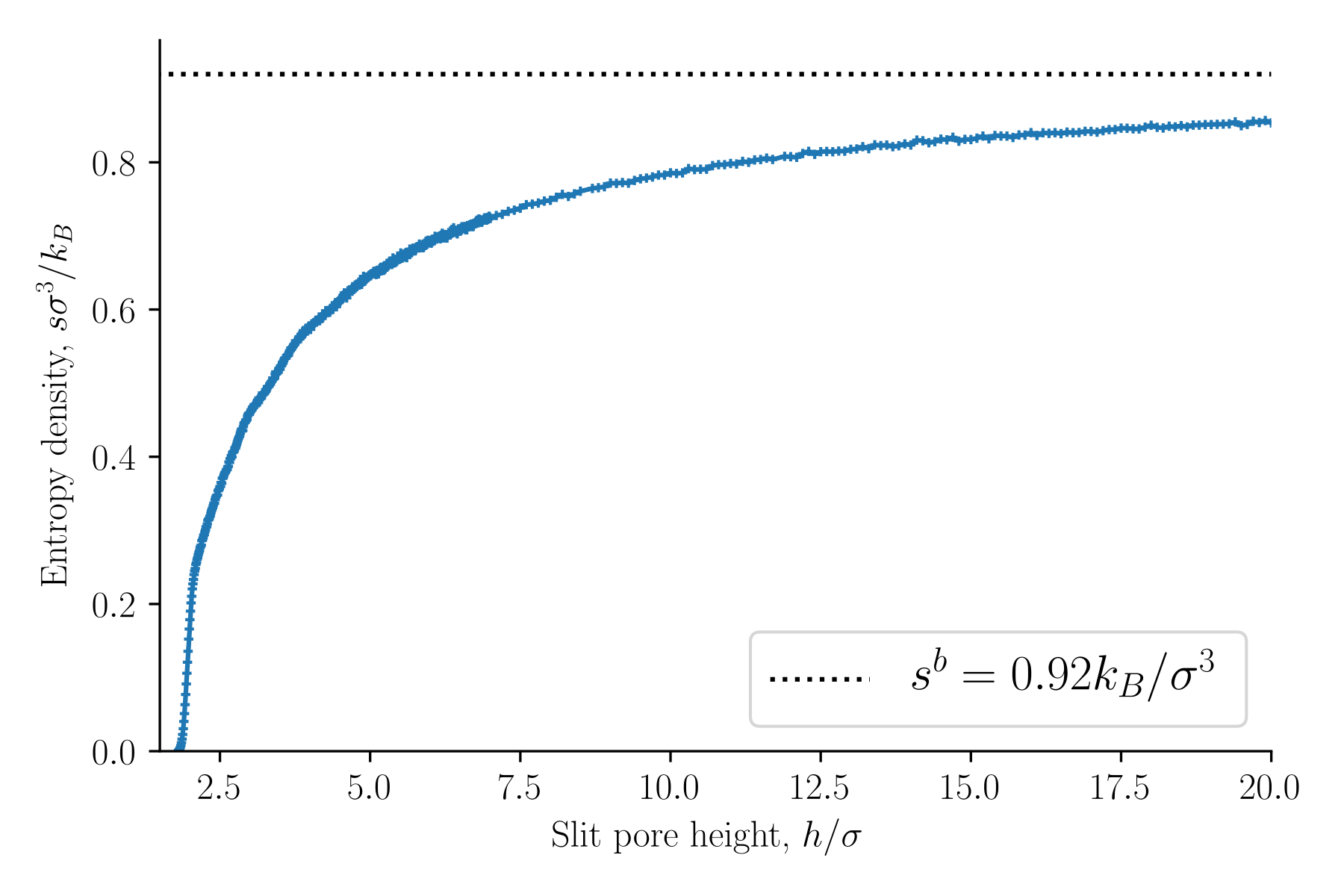}  
    \caption{Entropy density $s=S/V$ as a function of slit pore height $h$. See equation \ref{eq:entropy}. The dashed line shows the entropy density of the bulk $s^\text{b}$.} 
    \label{fig:entropy}
\end{figure}

\begin{figure}
    \centering
    \includegraphics[width=0.8\textwidth]{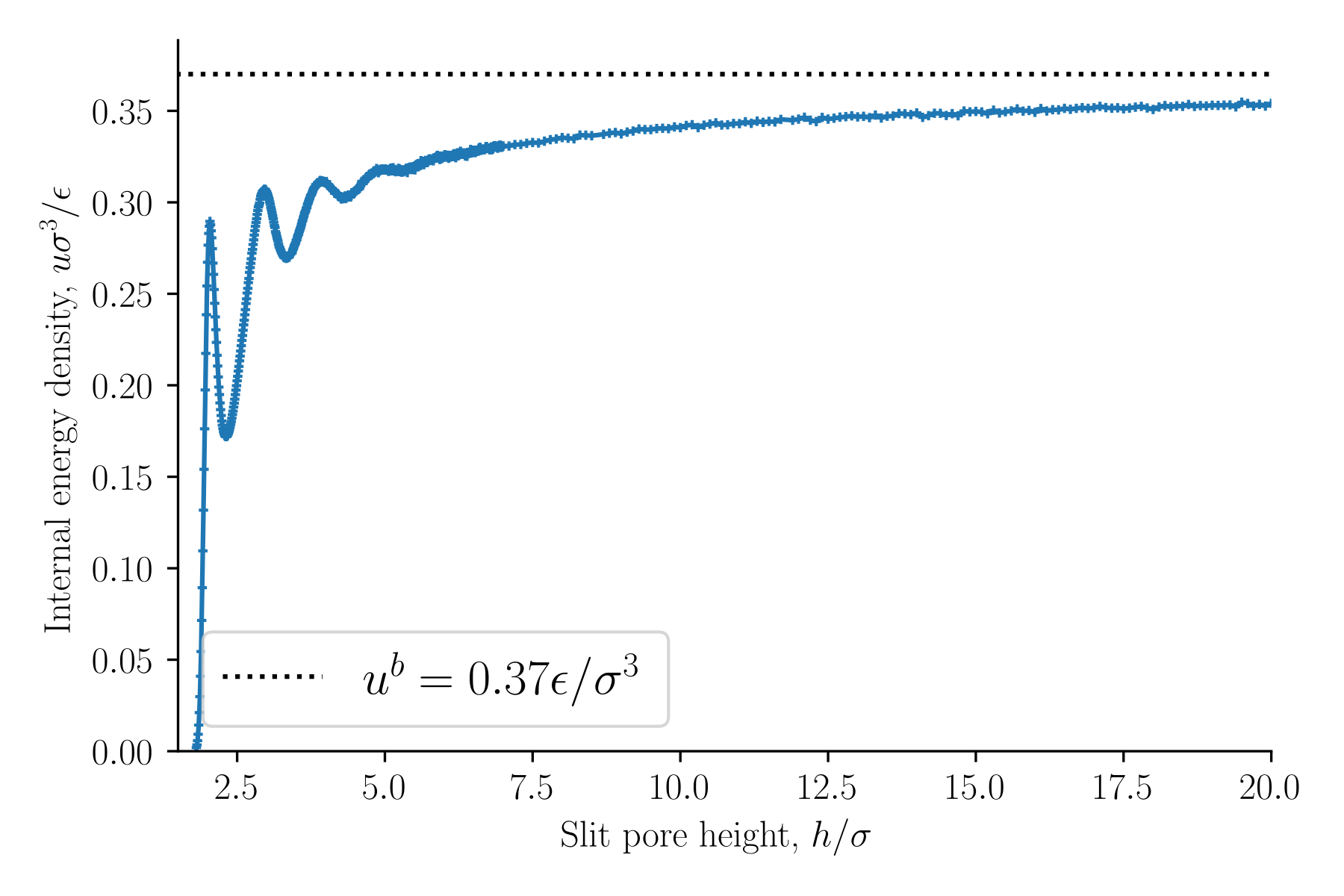}  
    \caption{Internal energy density $u=U/V$ as a function of slit pore height $h$. See equation \ref{eq:internal_energy3}. The dashed line shows the internal energy density of the bulk $u^\text{b}$.} 
    \label{fig:internal_energy}
\end{figure}

\begin{figure}
    \centering
    \includegraphics[width=0.8\textwidth]{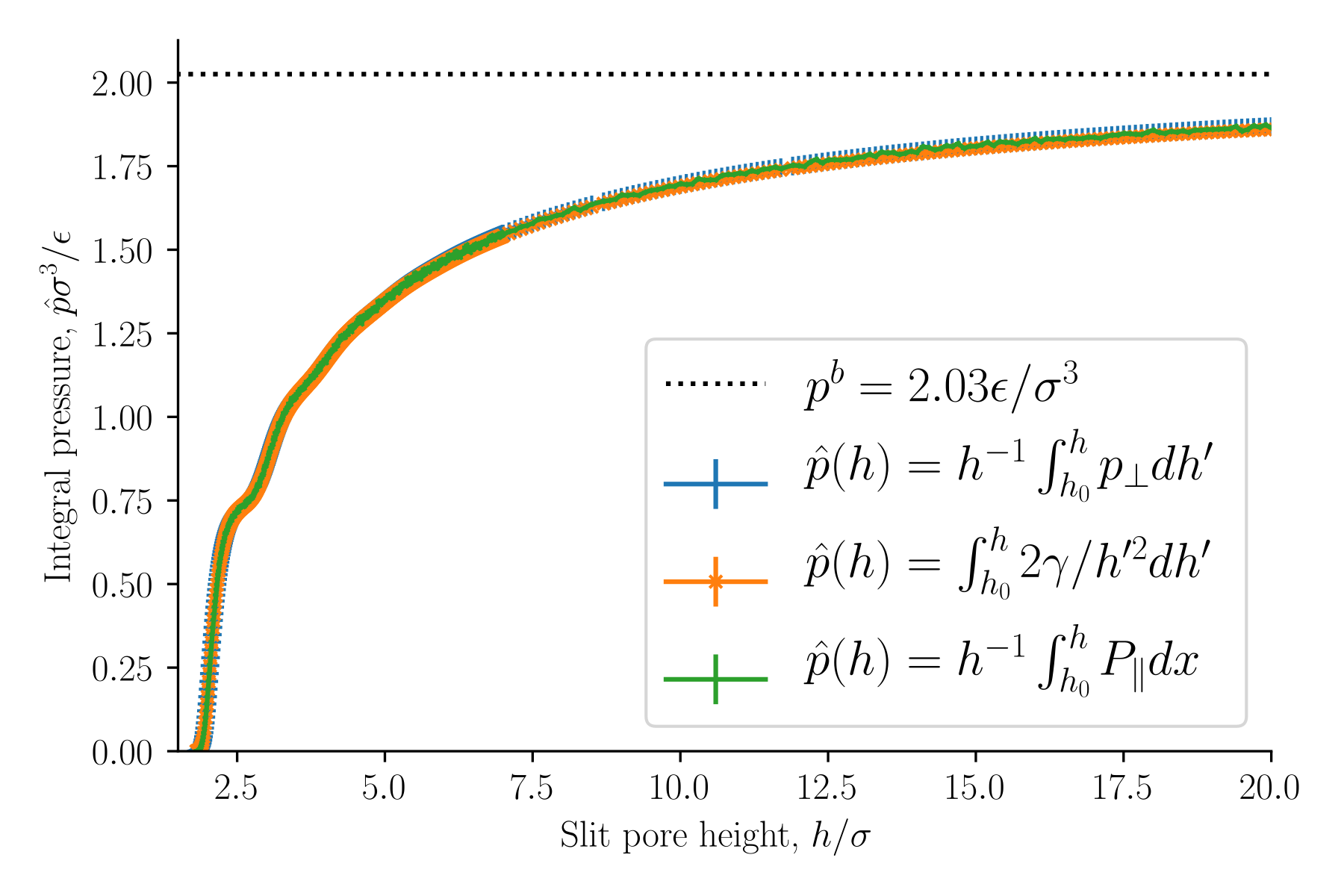}  
    \caption{Integral pressure $\hat{p}$ as a function of slit pore height $h$. The integral pressure is computed as the average tangential mechanical pressure, see equation \ref{eq:def_integral_pressure}, and from the normal pressure and surface tension, see equation \ref{eq:normal_pressure_and_surface_tension}.}
    \label{fig:integral_pressure}
\end{figure}

\begin{figure}
    \centering
    \includegraphics[width=0.8\textwidth]{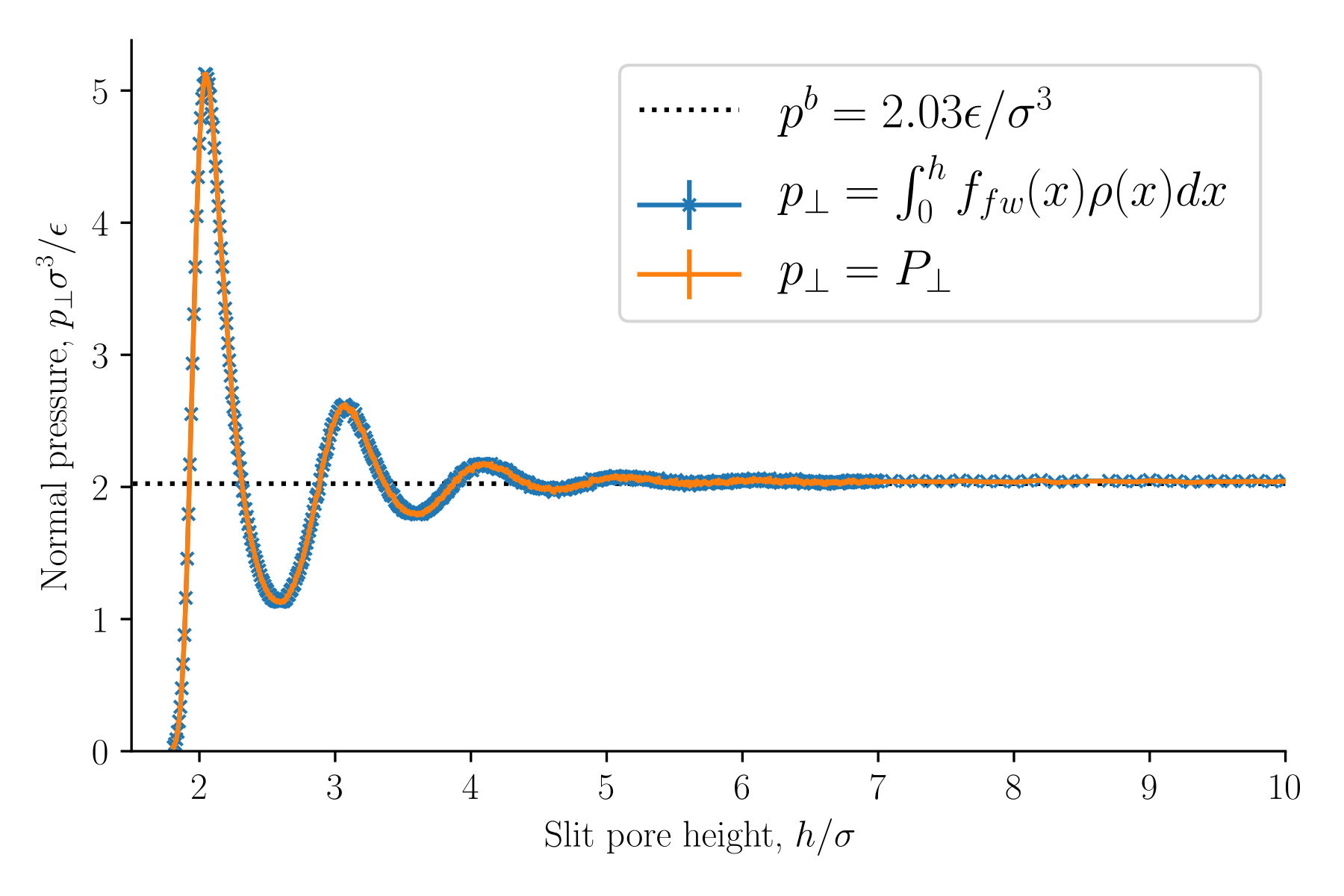}  
    \caption{Normal pressure $p_\perp =P_\perp $ as a function of slit pore
    height $h$. It is computed as the normal mechanical pressure tensor component, see equation \ref{eq:def_integral_normal_pressure}, and as the integral of the fluid-solid force times the local fluid density, see equation \ref{eq:disjoining_normal_density_profile}.} 
    \label{fig:normal_pressure}
\end{figure}

\begin{figure}
    \centering
    \includegraphics[width=0.8\textwidth]{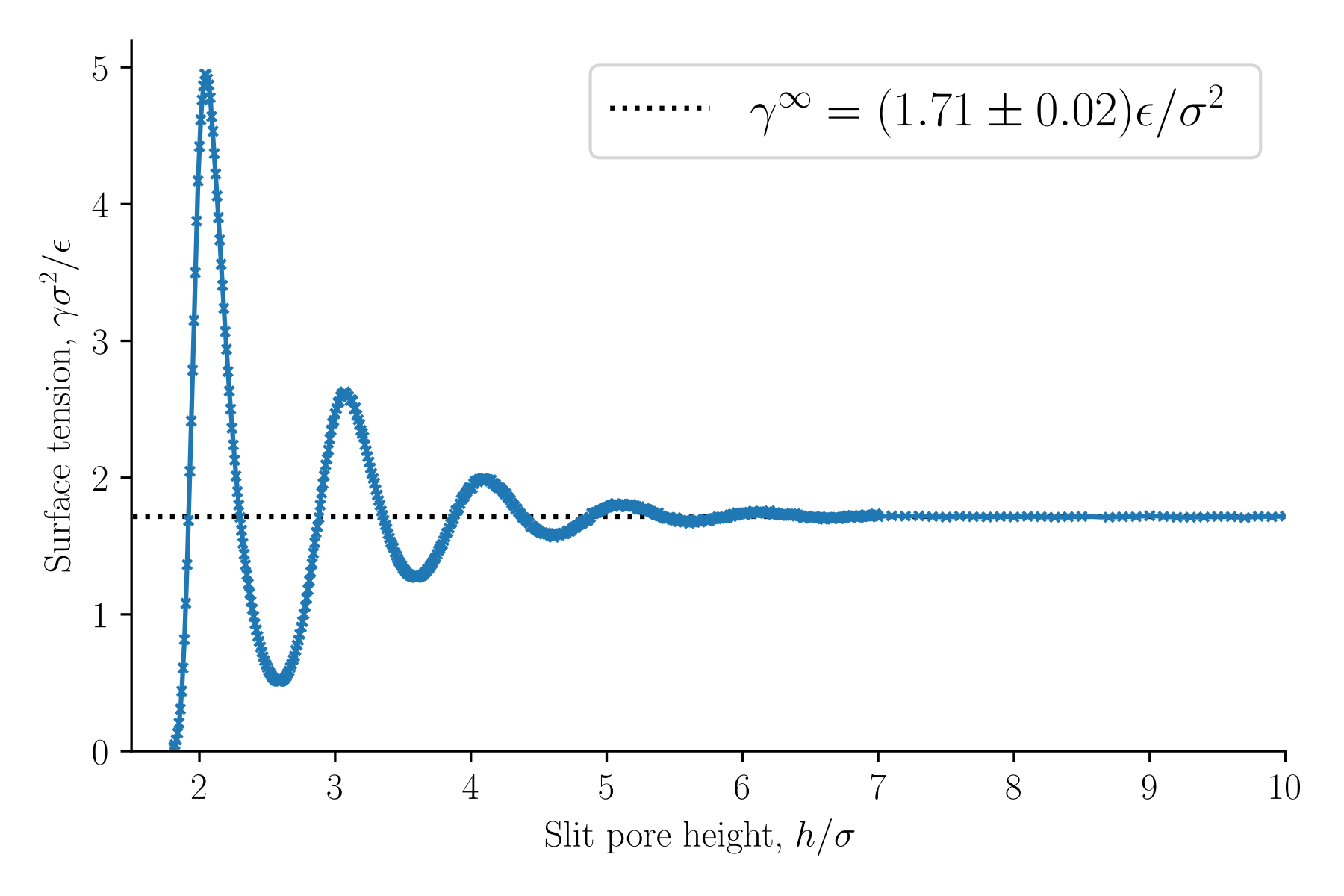}  
    \caption{Surface tension $\gamma$ as a function of slit pore height $h$, see equation \ref{eq:def_integral_surface_tension}. The dashed line shows the surface tension at infinite separation $\gamma^\infty$.}
    \label{fig:surface_tension}
\end{figure}

\begin{figure}
    \centering
    \includegraphics[width=0.8\textwidth]{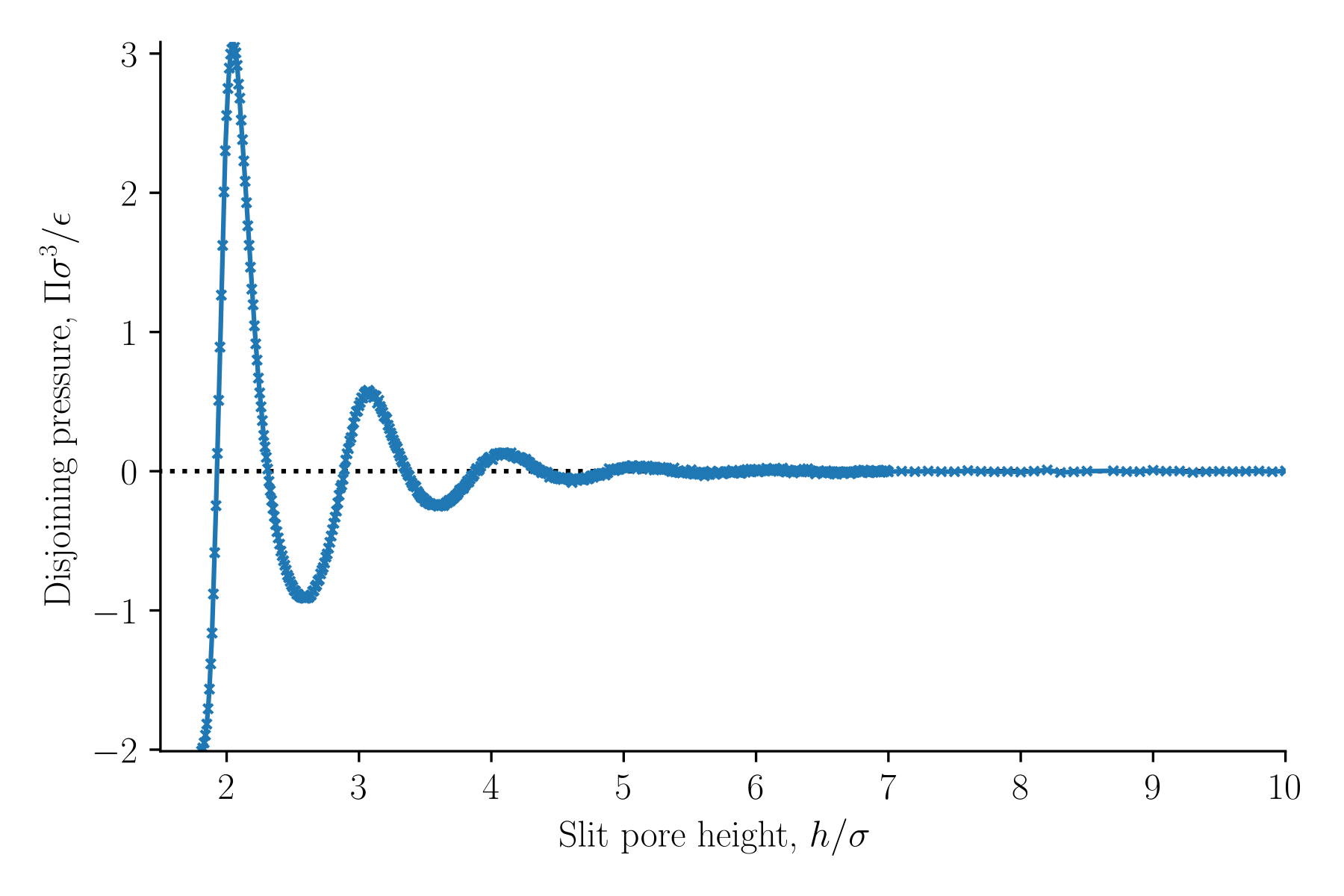}  
    \caption{Disjoining pressure as a function the slit pore height $h$, see equation \ref{eq:def_disjoining}.}
    \label{fig:disjoining_pressure}
\end{figure}

\begin{figure}
    \centering
    \includegraphics[width=0.8\textwidth]{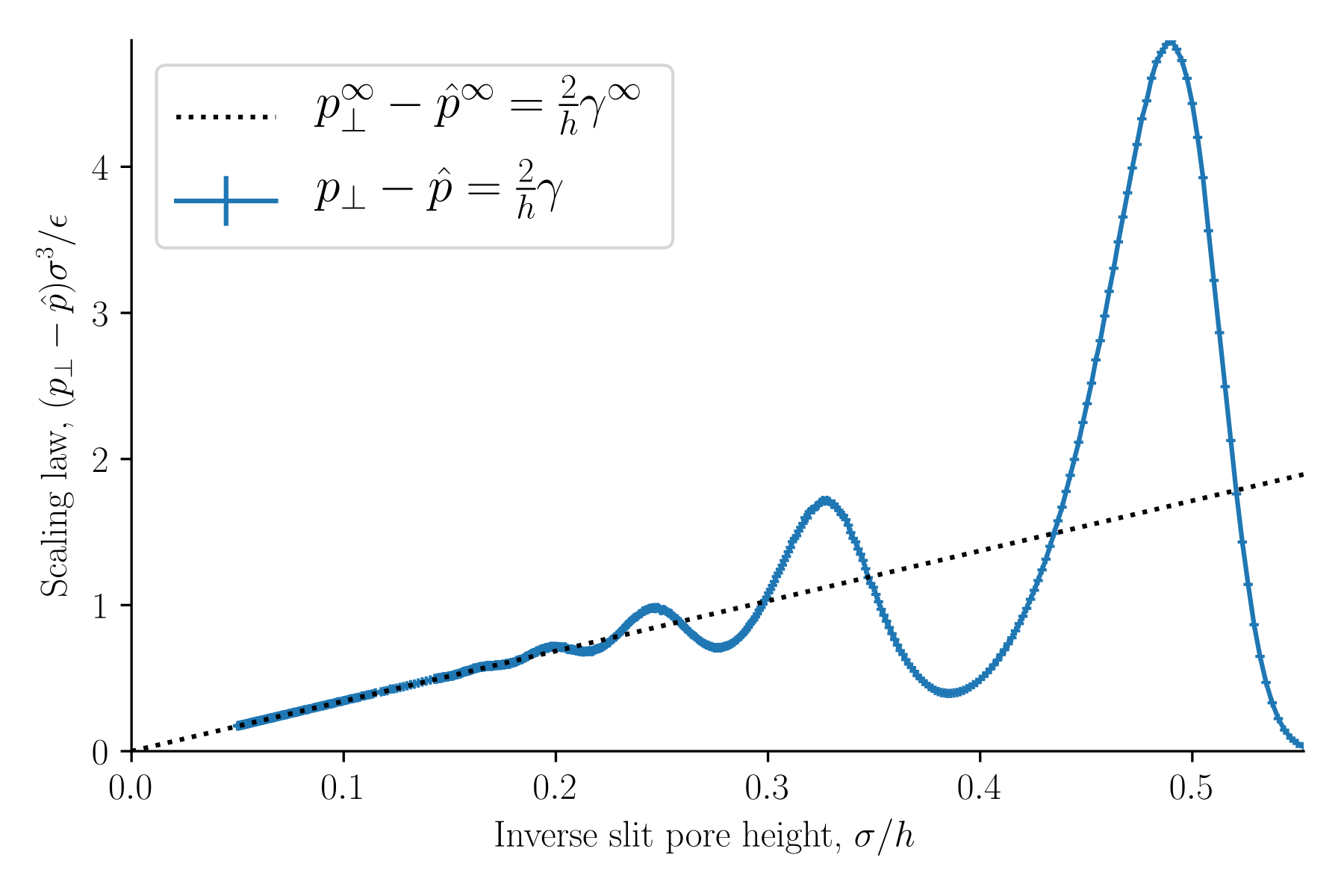}  
    \caption{The scaling of normal pressure minus integral pressure as a function of the inverse slit pore height $h$.}
    \label{fig:scaling_law}
\end{figure}

The results are presented in figures \ref{fig:profiles} to \ref{fig:scaling_law} and discussed in that order before general remarks are offered.

The normal mechanical pressure $P_\perp$ is presented in figure \ref{fig:profile_pn}. It does not depend on the position $x$, but it depends strongly on the slit pore height $h$. The figure shows a straight line of various lengths for each of the three heights, which reflect the slit pore height $h$. The integral normal pressure was identified as this component, $\hat{p}_\perp =P_\perp $. We see that it is always constant, as demanded by equation \ref{eq:def_integral_normal_pressure}. For slit pore heights $h>7\sigma$ we find that the normal mechanical pressure is equal to the bulk pressure. At $h=2.04\sigma$ the normal mechanical pressure is at a global maximum and at $h=2.59\sigma$ it has a local minimum for the given temperature and chemical potential. The normal mechanical pressures divided by the bulk pressure for the two cases are $P_\perp /p^{\text{b}}=2.5135\pm 0.0008$ and $P_\perp /p^{\text{b}}=0.5539\pm 0.0002$, respectively. The global minimum, which is zero, is at $h< 1.8\sigma$ when no fluid particles fit in the slit pore. This is because solid-solid interactions and quantum effects are not considered in this work. 

The tangential mechanical pressure $P_\parallel$, illustrated in figure \ref{fig:profile_pt}, depends in contrast on the position $x$ as well as on
the slit pore height $h$. The integral pressure is the average of $P_\parallel$, see equation \ref{eq:def_integral_pressure}. The tangential mechanical pressure follows the trend of the fluid number density, compare figures \ref{fig:profile_pt} and \ref{fig:profile_density}. For pore sizes $h>7\sigma$ the tangential mechanical pressure is constant and equal to the bulk pressure $p^{\text{b}}$ in the center of the pore. This indicates that the pore is large enough to accommodate bulk liquid in the center. The fluid is highly structured close to the fluid-solid surface \cite{Israelachvili1985}. As the slit pore height is decreased, fluid structures on the two sides overlap. When regions of structured fluids overlap, repulsive and attractive forces between the surfaces appear, and the disjoining pressure becomes non-zero. 

The fluid number density $\rho=N/V$ is presented in figure \ref{fig:density} as a function of the slit pore height $h$. The bulk fluid number density $\rho^\text{b}$ in equilibrium with the slit pore is shown as a dashed line. The fluid number density converges to the bulk value as the slit pore height approaches infinity. The volume $V$ depends on the choice of the fluid-solid dividing surfaces. We have chosen the dividing surfaces to be at $x=-L_x/2$ and $x=L_x/2$. The choice of the dividing surface determines how rapidly the slit pore values converges to the bulk values. A dividing surface closer to the fluid phase will reduce the volume and consequently the slit pore values will converge faster to the bulk values. Other choices of the dividing surface are possible. 

The entropy density is presented in figure \ref{fig:entropy} as a function of the slit pore height $h$. The entropy density is a monotonically increasing function of the height $h$. This confirms the observation by Israelachvili \cite{Israelachvili1985} that the origin of the oscillations of the disjoining pressure as a function of the height is not entropic. As a further confirmation of this point, we find that the internal energy density oscillates with a period equal to the particle diameter, see figure \ref{fig:internal_energy}. The oscillating forces or pressures are thus of energetic origin. The bulk entropy and internal energy densities are shown as dashed lines. The entropy and internal energy densities of the slit pore converges to the bulk values as the slit pore height is increased. 

The integral pressure is equal to the volume average tangential mechanical pressure $\hat{p}=h^{-1}\int_{0}^{h}P_{\parallel}\d x$. It is of special interest because it is equal to minus the grand potential divided by volume $\hat{p}=-X/V$, or the replica energy density. The grand potential is the starting point for the definition of the average thermodynamic properties of the REV \cite{Kjelstrup2018}. The integral pressure is presented in figure \ref{fig:integral_pressure} as a function of the slit pore height $h$. The integral pressure converges to the bulk pressure $p^\text{b}$ as the slit pore height approaches infinity as expected. The bulk pressure is shown as a dashed line. 

In previous works \cite{Kjelstrup2018, Kjelstrup2019, Galteland2019} we argued that the gradient of the integral pressure is the driving force for mass flux. In another work \cite{Rauter2020} we found the integral pressure of a two-phase system in a slit pore to be equal in the liquid and vapor in equilibrium. The identification of the integral pressure in this work is consistent with this interpretation. The gradient of the integral pressure is the driving force of the mass flux. For fluid flows tangential to the slit pore surfaces it is the gradient in the tangential mechanical pressure tensor component that gives the driving force when the system is out of equilibrium. In this work we identify the integral pressure as the average of the tangential mechanical pressure tensor components. 

As stated in section \ref{sec:Maxwell}, the integral and differential normal pressures and integral and differential surface tensions are expected to be equal when the surface area is large. If this is correct the integral pressure can be computed as  
\begin{equation}
	\hat{p}(h)=\frac{1}{h}\int_{h_{0}}^{h}\hat{p}_\perp h^{\prime}\quad =\int_{h_{0}}^{h}\frac{2\hat{\gamma}}{h^{\prime 2}}h^{\prime}.
	\label{eq:normal_pressure_and_surface_tension}
\end{equation}
The lower integration limit is $h_0=1.8\sigma$, at which point the integral pressure is in good approximation zero. The integral pressure computed from equation \ref{eq:normal_pressure_and_surface_tension} is shown in figure \ref{fig:integral_pressure}. The curves are identical. This implies that the integral and differential normal pressures are equal. As we have already shown that the subdivision potential is zero $\varepsilon=0$ for large surface areas with this set of control variables, it follows that the integral and differential surface tensions are also equal. We will from now on refer to the integral normal pressure and integral surface tensions as the normal pressure and surface tension. 

The normal pressure was identified as the normal mechanical pressure $p_\perp =P_\perp $ in equation \ref{eq:def_integral_normal_pressure}. It is presented in figure \ref{fig:normal_pressure} as a function of the slit pore height $h$. The normal pressure was also calculated from the local fluid density and the fluid-solid force using equation \ref{eq:disjoining_normal_density_profile}. The two methods of calculating the normal pressure agree, which indicates that we have calculated the mechanical pressure tensor correctly. 

The normal pressure oscillates with a period equal to the fluid particle diameter at small heights $h$. The oscillations decay as the height $h$ increases. Such oscillations have been observed in experiments and are well known, see for example Israelachvili \cite{Israelachvili1985}. The oscillations are caused by the structuring of the fluid particles between the surfaces, and by the fact that the fluid particles all have the same diameter. These oscillations are not present in a mixture of molecules with different diameters. As the height is increased above $h>7\sigma$ the oscillations vanish and the normal pressure is constant and equal to the bulk pressure. The bulk pressure is shown as a dashed line in the figure. At heights $h>7\sigma$ the fluid structuring near the walls do not overlap. At lower densities smaller oscillations are expected with faster decay. The normal pressure shows a similar trend to previous works \cite{Evans1986, Balbuena1993, Gubbins2014}. At very small heights the solid-solid interaction will dominate and completely overshadow the fluid-fluid and fluid-solid interactions presented here. We have not included any solid-solid interaction in this work, and as a consequence the normal pressure approaches zero because there is no room for any fluid particles to enter the slit pore. 

When $p_\perp =\hat{p}_\perp$ and $\varepsilon=0$, it follows from equation \ref{eq:epsilon} that the integral and differential surface tensions are equal $\gamma =\hat{\gamma}$. The surface tension as a function of the slit pore height is presented in figure \ref{fig:surface_tension}, see equation \ref{eq:def_integral_surface_tension}. The surface tension at infinite separation $\gamma^\infty$ is computed as the average surface tension of the slit pore, with height $h> 10\sigma $, at which point the surface tension is independent of $h$. 

The disjoining pressure $\Pi$ was computed from equation \ref{eq:def_disjoining}, and is shown as a function of the slit pore height $h$ in figure \ref{fig:disjoining_pressure}. The disjoining pressure was here defined to be equal to the integral normal pressure minus the normal pressure at infinite separation. Because the integral and the differential normal pressures are the same and because the normal pressure at infinite separation is equal to the bulk pressure in this case, the definition contain the commonly used definition of the disjoining pressure \cite{Derjaguin1934, Hansen1990}. The normal pressure is constant for slit pore heights $h>7\sigma$. Consequently, the normal pressure at infinite separation can be calculated as the normal pressure when $h>7\sigma$. The normal pressure at infinite separation is equal to the pressure in a bulk fluid with the same temperature and chemical potential. 

For the present case, we can claim that our definition of the disjoining pressure is equivalent to the common definition. Our definition is general, as it also covers the cases where the integral and the differential normal pressures are unequal. Examples where this is the case are given below. 

Figure \ref{fig:scaling_law} shows how the normal pressure minus the integral pressure scales with inverse slit pore height. It is equal to the scaling law presented in a previous work \cite{Rauter2020}. The slope of this ideal curve is equal to two times the surface tension. When the inverse slit pore height approaches zero, \textit{i.e.} when the walls are far apart, the normal pressure and integral pressures are equal as predicted. For pores that are so small that no bulk fluid can form in the center, the structuring at the walls starts to overlap, and a fluctuating difference is seen in the difference of the normal and integral pressures. In the region of the straight line, we have a scaling law, that relates states of different heights. Heights smaller than approximately $h<5\sigma$ the scaling law breaks down, the difference of the two pressures starts to oscillate. There are positive and negative deviations from the law. 

In our earlier work \cite{Rauter2020}, we studied liquid-vapor coexistence in a slit pore. In that work the surface area was not a control variable, and as a consequence the subdivision potential was found to be equal to  two times the surface tension divided by the slit pore height, see figure \ref{fig:scaling_law}. Another reasonable set of control variables is the height $h$ instead of the volume $V$. For a control variable set consisting of temperature, height, surface area and chemical potential, the differential normal pressure is equal to  
\begin{equation}
	p_\perp =\frac{1}{\Omega}\left(\frac{\partial(\hat{p}V)}{\partial h}\right)_{T,\Omega,\mu}=\hat{p}+h\left(\frac{\partial\hat{p}}{\partial h}\right)_{T,\Omega,\mu}.
\end{equation}
The differential surface tensions is   
\begin{equation}
	\gamma =-\frac{1}{2}\left(\frac{\partial(\hat{p}V)}{\partial\Omega}\right)_{T,\mu,h}=-\frac{\hat{p}h}{2}-\frac{V}{2}\left(\frac{\partial \hat{p}}{\partial\Omega}\right)_{T,\mu,h}.
\end{equation}
For this set of control variables the height $h$ is kept constant instead of the volume $V$. The subdivision potential is accordingly
\begin{equation}
    \begin{split}
        \varepsilon &=-\hat{p}V+p_{\perp }V+2\gamma \Omega\\
        &=\hat{p}V+hV\left( \frac{\partial \hat{p}}{\partial h}\right) _{T,\Omega ,\mu }+\Omega V\left( \frac{\partial \hat{p}}{\partial \Omega }\right) _{T,h,\mu }. 
    \end{split}
\end{equation}
These relations also help us characterize the smallness of the slit pore with the large walls. The subdivision potential, introduced by Hill as a measure of smallness, deviates from zero in the last relation, also when the integral pressure does not depend on the size of the area, with height $h$ and surface area $\Omega$ as control variables. A dependency on area is relevant when adsorption takes place on small spheres \cite{Strom2020}. A slit pore with large walls, may be expected to be small for small heights $h$, since the confined fluid is not bulk-like. However, we have seen that when we use volume $V$ and surface area $\Omega$ as control variables, the subdivision potential is zero for large surface areas. A zero subdivision potential means that the system also can be described perfectly using regular thermodynamics \cite{Gjennestad2020a}. It is nevertheless meaningful to define a non-zero integral pressure, because the integral pressure enters the grand potential. It therefore determines the thermodynamic properties of a REV. Away from equilibrium, it will create a driving force. 

The grand potential or minus the integral pressure times the volume are equal to the replica energy. The replica energy is not zero in the present case. Clearly, we have here an example where smallness is brought out in Hill's terms through the replica potential. 

\section{Conclusion}
\label{sec:conclusion}

We have developed a nanothermodynamic description based on the ideas of Hill to describe slit pores. As environmental control variables we chose the chemical potential, temperature, volume and surface area. We have seen that the outcome varies with the set chosen, but the procedure can be used for complex geometries and different sets. 

Following Hill, we introduced the subdivision potential. It is non-zero only when the integral pressure depends on the surface area $\Omega$. For large surface areas we find the subdivision potential is zero, and that $p_\perp =\hat{p}_\perp$ and $\gamma =\hat{\gamma}$. In this sense nanothermodynamics is equivalent to the usual thermodynamic description for all heights. The replica energy and therefore the integral pressure are non-zero. This allows us to identify a scaling law, which confirms earlier results \cite{Rauter2020, Erdos2020}. By choosing height $h$ rather than volume $V$ among the control variables, a non-zero subdivision potential appears. 

We have identified the thermodynamic properties by their mechanical counterparts in a consistent manner. The integral pressure, which is equal to minus the grand potential divided by volume, can be understood as the average tangential mechanical pressure. The normal pressure is the normal mechanical pressure, and the surface tension is the integral of the normal minus the tangential mechanical pressure. The entropy and internal energy densities vary with the slit pore height, confirming the observations of Israelachvili \cite{Israelachvili1985}. The entropy density increased monotonically with increasing height, while the energy density oscillates. This confirms that the disjoining pressure is not of entropic origin, it is of energetic origin \cite{Israelachvili1985}. 

By these investigations of the nanothermodynamic theory and the subsequent simulations, we hope to have expanded on the knowledge on Hill's method, making it more available for further studies, for instance of flow and reactions in porous media. 

\section{Acknowledgments}
Computer resources have been provided by Faculty of Natural Science at NTNU and by the HPC resources at UIT and NTNU provided by, www.sigma2.no. We are grateful to the Research Council of Norway for funding through its Center of Excellence funding scheme, project number 262644, PoreLab. 

\bibliographystyle{ieeetr}
\bibliography{library.bib}

\begin{thebibliography}{10}

\bibitem{Israelachvili1985}
J.~N. Israelachvili, {\em Intermolecular and surface forces}.
\newblock Academic press, 1985.

\bibitem{Mcdonald2015}
T.~M. McDonald, J.~A. Mason, X.~Kong, E.~D. Bloch, D.~Gygi, A.~Dani,
  V.~Crocella, F.~Giordanino, S.~O. Odoh, W.~S. Drisdell, {\em et~al.},
  ``Cooperative insertion of $\text{CO}_2$ in diamine-appended metal-organic
  frameworks,'' {\em Nature}, vol.~519, no.~7543, pp.~303--308, 2015.

\bibitem{Vlugt1999}
T.~Vlugt, R.~Krishna, and B.~Smit, ``Molecular simulations of adsorption
  isotherms for linear and branched alkanes and their mixtures in silicalite,''
  {\em J. Phys. Chem. B}, vol.~103, no.~7, pp.~1102--1118, 1999.

\bibitem{Bresme2007}
F.~Bresme and M.~Oettel, ``Nanoparticles at fluid interfaces,'' {\em J. Phys.
  Condens. Matter}, vol.~19, no.~41, p.~413101, 2007.

\bibitem{Bresme2009}
F.~Bresme, H.~Lehle, and M.~Oettel, ``Solvent-mediated interactions between
  nanoparticles at fluid interfaces,'' {\em J. Chem. Phys.}, vol.~130, no.~21,
  p.~214711, 2009.

\bibitem{Galteland2020}
O.~Galteland, F.~Bresme, and B.~Hafskjold, ``Solvent-mediated forces between
  ellipsoidal nanoparticles adsorbed at liquid–vapor interfaces,'' {\em
  Langmuir}, vol.~0, no.~0, p.~null, 0.
\newblock PMID: 33236631.

\bibitem{Derjaguin1934}
B.~Derjaguin, ``Untersuchungen {\"u}ber die reibung und adh{\"a}sion, iv,''
  {\em Kolloid Z.}, vol.~69, no.~2, pp.~155--164, 1934.

\bibitem{Moura2019}
M.~Moura, E.~G. Flekkøy, K.~J. Måløy, G.~Schäfer, and R.~Toussaint,
  ``Connectivity enhancement due to film flow in porous media,'' {\em Phys.
  Rev. Fluids}, vol.~4, no.~9, p.~094102, 2019.

\bibitem{Das2005}
D.~Das and S.~Hassanizadeh, {\em Upscaling multiphase flow in porous media}.
\newblock Springer, 2005.

\bibitem{Khanamiri2018}
H.~H. Khanamiri, C.~F. Berg, P.~A. Slotte, S.~Schl{\"u}ter, and
  O.~Tors{\ae}ter, ``Description of free energy for immiscible two-fluid flow
  in porous media by integral geometry and thermodynamics,'' {\em Water Resour.
  Res.}, vol.~54, no.~11, pp.~9045--9059, 2018.

\bibitem{Armstrong2019}
R.~T. Armstrong, J.~E. McClure, V.~Robins, Z.~Liu, C.~H. Arns, S.~Schl{\"u}ter,
  and S.~Berg, ``Porous media characterization using minkowski functionals:
  theories, applications and future directions,'' {\em Transp. Porous Med.},
  vol.~130, no.~1, pp.~305--335, 2019.

\bibitem{Slotte2020}
P.~A. Slotte, C.~F. Berg, and H.~H. Khanamiri, ``Predicting resistivity and
  permeability of porous media using minkowski functionals,'' {\em Transp.
  Porous Med.}, vol.~131, no.~2, pp.~705--722, 2020.

\bibitem{Kjelstrup2018}
S.~Kjelstrup, D.~Bedeaux, A.~Hansen, B.~Hafskjold, and O.~Galteland,
  ``Non-isothermal transport of multi-phase fluids in porous media. the entropy
  production,'' {\em Front. Phys.}, vol.~6, p.~126, 2018.

\bibitem{Kjelstrup2019}
S.~Kjelstrup, D.~Bedeaux, A.~Hansen, B.~Hafskjold, and O.~Galteland,
  ``Non-isothermal transport of multi-phase fluids in porous media.
  constitutive equations,'' {\em Front. Phys.}, vol.~6, p.~150, 2019.

\bibitem{Balbuena1993}
P.~B. Balbuena, D.~Berry, and K.~E. Gubbins, ``Solvation pressures for simple
  fluids in micropores,'' {\em J. Phys. Chem.}, vol.~97, no.~4, pp.~937--943,
  1993.

\bibitem{Gubbins2014}
K.~E. Gubbins, Y.~Long, and M.~{\'S}liwinska-Bartkowiak, ``Thermodynamics of
  confined nano-phases,'' {\em J. Chem. Thermodyn.}, vol.~74, pp.~169--183,
  2014.

\bibitem{Bedeaux2018}
D.~Bedeaux and S.~Kjelstrup, ``Hill’s nano-thermodynamics is equivalent with
  gibbs’ thermodynamics for surfaces of constant curvatures,'' {\em Chem.
  Phys. Lett.}, vol.~707, pp.~40--43, 2018.

\bibitem{Gjennestad2020a}
M.~A. Gjennestad and {\O}.~Wilhelmsen, ``Thermodynamic stability of volatile
  droplets and thin films governed by the disjoining pressure in open and
  closed containers,'' {\em Langmuir}, 2020.

\bibitem{Gjennestad2020b}
M.~A. Gjennestad and {\O}.~Wilhelmsen, ``Thermodynamic stability of droplets,
  bubbles and thick films in open and closed pores,'' {\em Fluid Phase
  Equilibr.}, vol.~505, p.~112351, 2020.

\bibitem{Strom2017}
B.~A. Str{\o}m, J.-M. Simon, S.~K. Schnell, S.~Kjelstrup, J.~He, and
  D.~Bedeaux, ``Size and shape effects on the thermodynamic properties of
  nanoscale volumes of water,'' {\em Phys. Chem. Chem. Phys.}, vol.~19, no.~13,
  pp.~9016--9027, 2017.

\bibitem{Galteland2019}
O.~Galteland, D.~Bedeaux, B.~Hafskjold, and S.~Kjelstrup, ``Pressures inside a
  nano-porous medium. the case of a single phase fluid,'' {\em Front. Phys.},
  vol.~7, p.~60, 2019.

\bibitem{Erdos2020}
M.~Erd{\H{o}}s, O.~Galteland, D.~Bedeaux, S.~Kjelstrup, O.~A. Moultos, and
  T.~J. Vlugt, ``Gibbs ensemble monte carlo simulation of fluids in
  confinement: Relation between the differential and integral pressures,'' {\em
  Nanomaterials}, vol.~10, no.~2, p.~293, 2020.

\bibitem{Rauter2020}
M.~T. Rauter, O.~Galteland, M.~Erd{\H{o}}s, O.~A. Moultos, T.~J. Vlugt, S.~K.
  Schnell, D.~Bedeaux, and S.~Kjelstrup, ``Two-phase equilibrium conditions in
  nanopores,'' {\em Nanomaterials}, vol.~10, no.~4, p.~608, 2020.

\bibitem{Strom2020}
B.~A. Str{\o}m, J.~He, D.~Bedeaux, and S.~Kjelstrup, ``When thermodynamic
  properties of adsorbed films depend on size: Fundamental theory and case
  study,'' {\em Nanomaterials}, vol.~10, no.~9, p.~1691, 2020.

\bibitem{Bedeaux2020}
D.~Bedeaux, S.~Kjelstrup, and S.~K. Schnell, {\em Nanothermodynamics. General
  theory}.
\newblock NTNU, Trondheim, Norway, 2020.

\bibitem{Hill1963}
T.~L. Hill, {\em Thermodynamics of small systems, part 1}.
\newblock Benjamin, 1963.

\bibitem{Hill1964}
T.~L. Hill, {\em Thermodynamics of small systems, part 2}.
\newblock Benjamin, 1964.

\bibitem{Hill1998}
T.~L. Hill and R.~V. Chamberlin, ``Extension of the thermodynamics of small
  systems to open metastable states: An example,'' {\em Proc. Natl. Acad. Sci.
  U.S.A}, vol.~95, no.~22, pp.~12779--12782, 1998.

\bibitem{Hill2002}
T.~L. Hill and R.~V. Chamberlin, ``Fluctuations in energy in completely open
  small systems,'' {\em Nano Lett.}, vol.~2, no.~6, pp.~609--613, 2002.

\bibitem{Hansen1990}
J.-P. Hansen and I.~R. McDonald, {\em Theory of simple liquids}.
\newblock Elsevier, 1990.

\bibitem{Radke2015}
C.~Radke, ``Film and membrane-model thermodynamics of free thin liquid films,''
  {\em J. Colloid Interf. Sci.}, vol.~449, pp.~462--479, 2015.

\bibitem{Long2011}
Y.~Long, J.~C. Palmer, B.~Coasne, M.~{\'S}liwinska-Bartkowiak, and K.~E.
  Gubbins, ``Pressure enhancement in carbon nanopores: a major confinement
  effect,'' {\em Phys. Chem. Chem. Phys.}, vol.~13, no.~38, pp.~17163--17170,
  2011.

\bibitem{Dijk2020}
D.~van Dijk, ``Comment on "pressure enhancement in carbon nanopores: a major
  confinement effect" by y. long, j. c. palmer, b. coasne, m.
  sliwinska-bartkowiak and k. e. gubbins, phys. chem. chem. phys., 2011, 13,
  17163,'' {\em Phys. Chem. Chem. Phys.}, vol.~22, no.~17, pp.~9824--9825,
  2020.

\bibitem{Long2020}
Y.~Long, J.~C. Palmer, B.~Coasne, K.~Shi, M.~{\'S}liwi{\'n}ska-Bartkowiak, and
  K.~E. Gubbins, ``Reply to the ‘comment on “pressure enhancement in carbon
  nanopores: a major confinement effect”’by d. van dijk, phys. chem. chem.
  phys., 2020, 22,'' {\em Phys. Chem. Chem. Phys.}, vol.~22, no.~17,
  pp.~9826--9830, 2020.

\bibitem{Schofield1982}
P.~Schofield and J.~R. Henderson, ``Statistical mechanics of inhomogeneous
  fluids,'' {\em Proc. R. Soc. Lon. Ser.-A}, vol.~379, no.~1776, pp.~231--246,
  1982.

\bibitem{Harasima1958}
A.~Harasima, ``Molecular theory of surface tension,'' {\em Adv. Chem. Phys},
  vol.~1, pp.~203--237, 1958.

\bibitem{Irving1950}
J.~Irving and J.~G. Kirkwood, ``The statistical mechanical theory of transport
  processes. iv. the equations of hydrodynamics,'' {\em J. Chem. Phys.},
  vol.~18, no.~6, pp.~817--829, 1950.

\bibitem{Hafskjold2002}
B.~Hafskjold and T.~Ikeshoji, ``Microscopic pressure tensor for hard-sphere
  fluids,'' {\em Phys. Rev. E}, vol.~66, no.~1, p.~011203, 2002.

\bibitem{Shi2020}
K.~Shi, Y.~Shen, E.~E. Santiso, and K.~E. Gubbins, ``Microscopic pressure
  tensor in cylindrical geometry: Pressure of water in a carbon nanotube,''
  {\em J. Chem. Theory Comput.}, vol.~16, no.~9, pp.~5548--5561, 2020.

\bibitem{Ikeshoji2003}
T.~Ikeshoji, B.~Hafskjold, and H.~Furuholt, ``Molecular-level calculation
  scheme for pressure in inhomogeneous systems of flat and spherical layers,''
  {\em Mol. Simulat.}, vol.~29, no.~2, pp.~101--109, 2003.

\bibitem{Evans1986}
R.~Evans, U.~M.~B. Marconi, and P.~Tarazona, ``Fluids in narrow pores:
  Adsorption, capillary condensation, and critical points,'' {\em J. Chem.
  Phys.}, vol.~84, no.~4, pp.~2376--2399, 1986.

\bibitem{Frenkel2001}
D.~Frenkel and B.~Smit, {\em Understanding Molecular Simulation: From
  Algorithms to Applications}, vol.~1.
\newblock Elsevier, 2001.

\bibitem{Shinoda2004}
W.~Shinoda, M.~Shiga, and M.~Mikami, ``Rapid estimation of elastic constants by
  molecular dynamics simulation under constant stress,'' {\em Phys. Rev. B},
  vol.~69, no.~13, p.~134103, 2004.

\bibitem{Plimpton1995}
S.~Plimpton, ``Fast parallel algorithms for short-range molecular dynamics,''
  {\em J. Comput. Phys.}, vol.~117, no.~1, pp.~1--19, 1995.

\bibitem{Hafskjold2019}
B.~Hafskjold, K.~P. Travis, A.~B. Hass, M.~Hammer, A.~Aasen, and
  {\O}.~Wilhelmsen, ``Thermodynamic properties of the 3d lennard-jones/spline
  model,'' {\em Mol. Phys.}, vol.~117, no.~23-24, pp.~3754--3769, 2019.

\bibitem{Stukowski2009}
A.~Stukowski, ``Visualization and analysis of atomistic simulation data with
  ovito--the open visualization tool,'' {\em Model. Simul. Mater. Sc.},
  vol.~18, no.~1, p.~015012, 2009.

\end{thebibliography}

\end{document}